\documentclass[10pt,twocolumn,aps,prd,showpacs,superscriptaddress,nofootinbib,amsmath,amssymb,floats,floatfix,showkeys,notitlepage,longbibliography]{revtex4-2}
\usepackage{orcidlink}
\usepackage{comment}
\usepackage{lipsum}
\usepackage{graphicx}
\usepackage{subfigure}
\usepackage{palatino}
\usepackage{sans}
\usepackage{hyperref}
\hypersetup{colorlinks=true,linkcolor=blue,urlcolor=blue,citecolor=blue}
\usepackage[toc,page]{appendix}
\usepackage[normalem]{ulem}
\usepackage{adjustbox}
\usepackage{latexsym}
\usepackage{amsmath}
\usepackage{amssymb}
\usepackage{amsfonts}
\usepackage{dcolumn}
\usepackage{bm}
\usepackage{tikz}
\usetikzlibrary{decorations.pathmorphing}
\usepackage{bigints}
\usepackage{array,tabularx,multirow,booktabs}
\usepackage[tracking=true]{microtype}
\usepackage{soul} 
\SetTracking{}{500}
\SetTracking{encoding={*}, shape=sc}{40}
\UseRawInputEncoding 
\allowdisplaybreaks

\usepackage[utf8]{inputenc}
\usepackage{xcolor} 

\renewcommand{\Im}{\ensuremath{\mathrm{Im}}}

\begin{document} \sloppy

\title{Josephson's effect in the Schwarzschild background}

\author{Reggie C. Pantig \orcidlink{0000-0002-3101-8591}} 
\email{rcpantig@mapua.edu.ph}
\affiliation{Physics Department, School of Foundational Studies and Education, Map\'ua University, 658 Muralla St., Intramuros, Manila 1002, Philippines.}

\author{Ali \"Ovg\"un \orcidlink{0000-0002-9889-342X}}
\email{ali.ovgun@emu.edu.tr}
\affiliation{Physics Department, Faculty of Arts and Sciences, Eastern Mediterranean University, Famagusta, 99628 North Cyprus via Mersin 10, Turkiye.}

\begin{abstract}
We develop a fully covariant, analytic framework for Josephson phenomena in static curved spacetimes and specialize it to the Schwarzschild exterior. The formulation rests on two invariant elements: the gauge-invariant condensate momentum that governs phase dynamics and the conserved current whose hypersurface flux encodes transport for an observer at infinity. Using the timelike Killing field to relate proper and asymptotic quantities, we derive a redshifted AC Josephson law in which the asymptotic phase-evolution rate is proportional to the difference of redshifted voltage drops, i.e. to \(V_i^\infty\equiv \alpha_i V_i^{\rm proper}\); equivalently, it depends on \(\alpha_i V_i^{\rm proper}\) for local control. Under RF drive specified at infinity, the Shapiro-step loci are invariant (expressed in asymptotic voltages) while propagation phases set any apparent lobe translation. For DC transport, a short-junction solution on a static slice yields the proper current-phase relation; mapping to asymptotic observables gives a single-power redshift scaling of critical currents, \(I_{c,\infty}\propto \alpha I_c^{\rm proper}\), whereas power scales as \(P_\infty\propto \alpha^2 P_{\rm proper}\). In a "vertical" dc-SQUID with junctions at different radii, gravity does not shift the DC interference pattern at linear order; it produces a small envelope deformation and an amplitude rescaling. Gravity does not alter the local Josephson microphysics; it reshapes the clocks and energy accounting that define measurements at infinity. The resulting predictions are gauge- and coordinate-invariant, operationally stated in terms an experimenter can control (proper vs. asymptotic bias), and remain analytic from the weak-field regime to the near-horizon limit.
\end{abstract}


\maketitle

\section{Introduction} \label{sec1}
Suppose we fabricate a Josephson weak link whose two superconducting banks sit at different gravitational potentials-one deeper in a static field (e.g., near a horizon) or on opposite ends of a uniformly accelerated platform. How does spacetime redshift reorganize the AC Josephson frequency-voltage relation, the DC critical current, and the interference pattern of a SQUID that straddles this potential difference? This question defines a program on gravito-Josephson effects: whether the measurable phase rate tracks redshifted voltages, whether the critical current seen by an asymptotic observer scales with a single power of the lapse while power scales with two, and whether SQUID lobes are shifted by gravity itself or only by propagation (geometric+Shapiro) phases. While the strongest manifestations are expected in compact-object environments, the same kinematics suggests weak-field signatures and metrology protocols that test the equivalence principle with superconducting circuits.

We study phase-coherent quantum transport in a static black-hole background using only analytic, covariant methods. We aim to recast the Josephson effects in a language that remains well defined for observers who do not share the proper times of the superconducting banks that form a weak link. The central structural point is that all measurable Josephson phenomena are statements about gauge-invariant phase differences and conserved currents. Foundational demonstrations of AC and DC Josephson physics and microwave locking include the original prediction \cite{Josephson:1962zz}, the observation of Shapiro steps \cite{Shapiro:1963nhj}, early microwave-induced dc voltages \cite{Langenberg_1966}, and textbook developments \cite{tinkham1975introduction}; the London equations remain the minimal local framework for superconducting electrodynamics \cite{London1935}. In a curved, static spacetime, these objects are naturally tied to the timelike Killing field, the Tolman-Ehrenfest redshift of intensive variables, and hypersurface fluxes that encode how local transport appears to an observer at infinity. Specializing to the Schwarzschild exterior allows us to make these ideas fully explicit while avoiding complications associated with rotation or time dependence. The conceptual backdrop for our construction is the Tolman-Ehrenfest redshift of intensive variables in static spacetimes, which fixes how local clocks and electrochemical potentials compare across different radii \cite{Tolman:1930zza,Tolman:1930ona}.  Early ideas that gravity can influence superconducting phases and supercurrents go back at least to DeWitt’s analysis of gravitational coupling to superconductors \cite{DeWitt:1966yi}. Related covariant and interferometric viewpoints include $3+1$ electrodynamics in curved spacetimes \cite{Thorne_1982}, hydrodynamics with a spontaneously broken $U(1)$ in relativistic media \cite{Son:2000ht}, and reviews on the interplay between superconductors and gravity \cite{Gallerati:2022pgh}.

Throughout, we restrict attention to static configurations of the background geometry, and we treat the electromagnetic sector as a classical test field. Dissipation, nonequilibrium noise, and microscopic pairing mechanisms are not modeled. Our results concern three regimes that are distinct in flat space and remain distinct here: the AC Josephson law and its frequency metrology, DC weak-link transport, and the scaling of the asymptotically measured critical current, and interferometry in a loop that spans different gravitational potentials. The thread that ties these regimes together is a simple dictionary between quantities defined locally in the proper frames of the banks and the corresponding fluxes and frequencies defined with respect to the Killing time. In Earth’s weak field, the frequency shift is tiny, in line with dedicated AC-Josephson estimates \cite{Ummarino:2020loo} and precision redshift tests with clocks and atom interferometers \cite{Pumpo:2021sok,Zheng:2022hwj}.

We work in geometric units unless noted otherwise. The Schwarzschild line element outside the horizon takes the standard form
\begin{equation} \label{metric}
ds^2 = -\alpha(r)^2dt^2 + \alpha(r)^{-2}dr^2 + r^2 d\Omega^2,
\end{equation}
where
\begin{equation} \label{alpha}
    \alpha(r)=\sqrt{1-\frac{2GM}{rc^2}},
\end{equation}
and the static Killing field is \(\xi^\mu=(\partial_t)^\mu\). The redshift factor \(\alpha(r)\) is the norm of \(\xi^\mu\) (with \(c=1\)). We will use \(\alpha\) to translate between proper-time rates and asymptotic frequencies, and we will use conserved Noether currents to translate between proper currents and asymptotically measured fluxes. These two translations are sufficient to reconstruct the full Josephson phenomenology in Schwarzschild spacetime.

In flat space, the Josephson relations express the dynamics of the gauge-invariant phase difference \(\Delta\varphi\) across a weak link in terms of electromagnetic control parameters. The DC relation yields a current-phase law \(I=I_c \sin\Delta\varphi\) for a short junction in the tunneling regime. The AC relation identifies the time derivative of \(\Delta\varphi\) with the electrochemical potential difference, leading to frequency-voltage metrology of the form \(\dot{\Delta\varphi}=(2e/\hbar)V\). Under periodic driving, one obtains phase locking and Shapiro steps. Historically, the DC and AC Josephson relations were predicted in flat space by Josephson \cite{Josephson:1962zz}, establishing the phase-voltage metrology that we generalize covariantly here. The separation between local constitutive physics and global redshift is complementary to gravitational/holographic Josephson realizations, which reproduce the current-phase relation in curved backgrounds \cite{Horowitz:2011dz}.

\begin{figure}
    \centering
\includegraphics[width=\columnwidth]{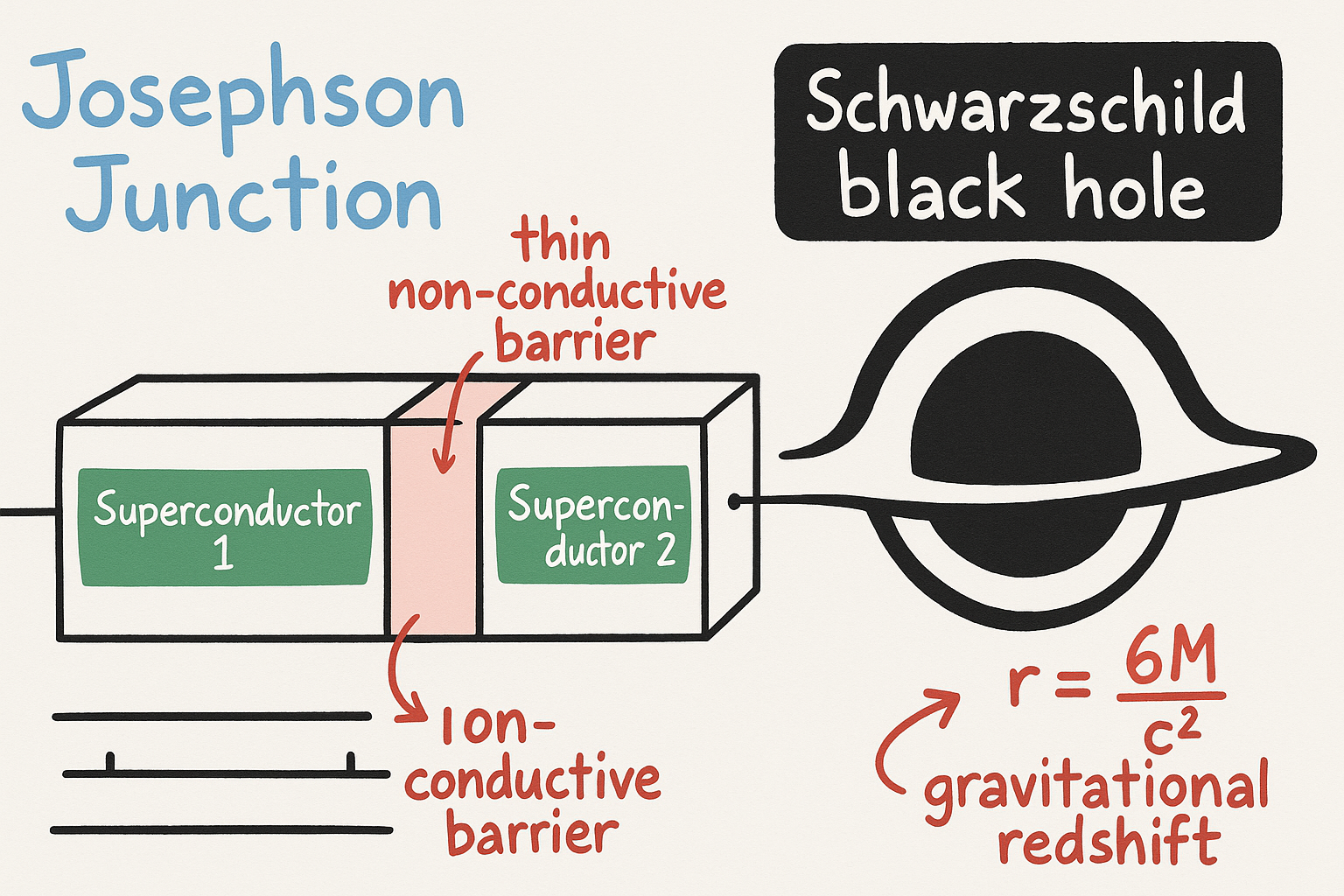}
    \caption{Schematic of a Josephson junction placed just outside a Schwarzschild event horizon.}
    \label{fig0}
\end{figure}

A curved, static background complicates two premises of this flat-space picture. First, the two banks may reside at different gravitational potentials, so their proper times do not agree. Second, potentials and temperatures redshift according to the Tolman-Ehrenfest laws. An asymptotic observer who uses the Killing time \(t\) does not directly measure the proper dynamics of either bank. Nevertheless, the Josephson relations remain statements about two invariant structures: the phase couples to electromagnetism through \(\partial_\mu\theta - q A_\mu/\hbar\), and transport is encoded in a conserved current \(j^\mu\). In a static geometry, these quantities admit a clean projection onto the space of Killing orbits. Related covariant perspectives on phase coherence in gravitational settings include relativistic treatments of interferometric loops, where inertial and gravitational effects imprint directly on gauge-invariant phases \cite{Anandan:1981zd,Anandan_1984}. General-relativistic extensions of electromagnetic transport in conductors and superconductors (including Ohm’s law in curved spacetime, rotating-frame corrections, and thermoelectric/gravito-electromagnetic effects) have been developed in a series of works \cite{Ahmedov:1998mif,Ahmedov:1999bqr,Ahmedov:2010mt,Das:2023qqx,Gavassino:2025bnx}.

Our first objective is to write the AC Josephson relation in a form that is manifestly gauge invariant and that cleanly separates proper-time dynamics from asymptotic measurement. The asymptotic phase-evolution rate \(\dot{\Delta\varphi}_\infty\) depends on the redshifted voltage drops: it is proportional to \(V_1^\infty - V_2^\infty\), where \(V_i^\infty\equiv \alpha_i V_i^{\rm proper}\), or, equivalently, to \(\alpha_1 V_1^{\rm proper}-\alpha_2 V_2^{\rm proper}\). Equal proper biases therefore do not imply equal asymptotic frequencies when \(\alpha_1\neq \alpha_2\). In the weak-field limit, this reduces to a small, controllable correction; near the horizon \(\alpha\to 0\) reorganizes the spectrum in a simple scaling form, while the step loci expressed in asymptotic voltages remain fixed when both bias and drive are specified at infinity. In the weak-field limit, the resulting frequency shifts are exceedingly small for terrestrial fields, consistent with estimates of gravitationally induced Josephson oscillations in Earth’s potential \cite{Ummarino:2020loo}.

Our second objective is to relate the critical current that governs a local weak link to the current measured at infinity. A short-junction solution on a static slice yields a proper current-phase relation. Mapping to asymptotic observables introduces one power of the lapse in the current, \(I_{c,\infty}\simeq \alpha I_c^{\rm proper}\), with curvature corrections controlled by junction size. By contrast, power picks up two powers, \(P_\infty=\alpha^2 P_{\rm proper}\). These are kinematic statements independent of barrier microphysics. This separation between local microphysics and global redshift is complementary to holographic constructions where curved spacetimes host Josephson junctions on the gravity side yet still reproduce the standard current-phase phenomenology \cite{Horowitz:2011dz}.

Our third objective is to analyze SQUID interferometry \cite{Jaklevic:1964ysq} in a loop whose arms lie at different radii. Flux quantization on a static slice has its familiar form and is insensitive to \(\alpha\). Consequently, in a vertical SQUID, the two junction phases can evolve at different rates for equal proper biases, but the DC interference pattern in external flux is not translated by gravity at linear order; instead, the envelope is slightly deformed, and the overall amplitude rescales. Under RF drive from infinity, any apparent lobe translation is governed by differential propagation phases (geometric plus Shapiro). Redshift alone does not shift the step loci when both the bias and the drive are specified at infinity. Closely related interferometric effects have been analyzed in relativistic settings, where gravitational and inertial phases enter loop constraints; our treatment isolates the static redshift piece and clarifies when apparent lobe translations arise from propagation phases rather than redshift itself \cite{Anandan:1981zd,Anandan_1984}. Interference shifts under RF drive are further constrained by propagation phases, including geometric path differences and Shapiro time delay \cite{Shapiro:1964uw,Possel:2019zbu}, while gravitational Aharonov-Bohm-type phases have been analyzed and observed in atomic platforms \cite{Overstreet:2021hea,Chiao:2023ezj}. Also, a kind of Josephson junction is formed in the vicinity of the throat of a traversable wormhole \cite{Takeuchi:2021kjv}.

Conceptually, we emphasize two methodological choices. First, we formulate everything on static hypersurfaces with the induced metric \(\gamma_{ij}\). All transport statements are then expressed as hypersurface fluxes constructed from \(j^\mu\). Second, we use the Killing symmetry to define the quantities that an asymptotic observer would measure. This is natural in Schwarzschild and keeps the analysis coordinate independent. No numerical evolution is required.

The paper establishes three main results, each derived in closed form for the Schwarzschild exterior. (i) The redshifted AC Josephson law: for banks at radii \(r_1\) and \(r_2\) with \(\alpha_i=\alpha(r_i)\), the asymptotic phase rate obeys
\begin{equation}
    \dot{\Delta\varphi}_\infty=\frac{2e}{\hbar} (V_1^\infty-V_2^\infty)=\frac{2e}{\hbar} \left(\alpha_1 V_1^{\rm proper}-\alpha_2 V_2^{\rm proper}\right).
\end{equation}
(ii) DC transport and the single-\(\alpha\) law: for a short junction on a static slice, the asymptotic critical current scales as \(I_{c,\infty}\simeq \alpha I_c^{\rm proper}\), while power scales as \(\alpha^2\). (iii) Interference in a vertical SQUID: flux quantization supplies the loop constraint; phase evolution carries the appropriate \(\alpha_i\); in DC operation, the lobe centers are not shifted at linear order, and in RF operation, any lobe translation tracks differential propagation phases rather than redshift. These statements are framed so that both the flat-space limits and the near-horizon scalings are transparent.

We organize the paper as follows: Section \ref{sec2} develops the covariant kinematics of superconducting phases and currents in static spacetimes and records the proper-to-asymptotic conversion rules used throughout. Section \ref{sec3} derives the redshifted AC Josephson law and analyzes phase locking under periodic driving from infinity, including weak-field expansions and near-horizon scaling. Section \ref{sec4} treats DC weak-link transport on a Schwarzschild static slice and establishes the single-\(\alpha\) scaling of the asymptotic critical current with curvature corrections. Section \ref{sec5} analyzes a vertical SQUID whose junctions are at different radii and presents closed-form expressions for the interference envelope and RF phase translations. The discussion in Section \ref{sec6} summarizes the conceptual lessons, clarifies the domain of validity, and outlines extensions to slowly rotating backgrounds and analogue platforms. We work under the following standing assumptions: (i) static background geometry with timelike Killing field \(\xi^\mu\) and zero shift; (ii) short-junction limit across the barrier, \(L\ll R_c\) and \(L\ll \ell_\alpha\), so that metric/lapse variations across the barrier are negligible; (iii) classical electromagnetism on the fixed background (test-field limit); (iv) banks comoving with static observers; (v) all “proper” quantities live on a static slice \(\Sigma_t\), whereas “asymptotic” observables are fluxes and frequencies with respect to Killing time; (vi) when converting between proper and asymptotic observables over the junction area we may take \(\alpha\) uniform, though different banks may sit at different \(\alpha_{1,2}\).
Beyond black-hole exteriors, the same invariant dictionary may inform precision tests of gravity with phase-coherent systems, dovetailing with recent proposals for gravitational metrology that exploit Josephson-like oscillations as frequency standards \cite{Lammerzahl2024}.

\section{Relativistic kinematics of superflow in static spacetimes} \label{sec2}
We now develop the covariant framework that underpins all subsequent calculations. The setting is a static spacetime $(\mathcal{M},g_{\mu\nu})$ with a timelike Killing field $\xi^\mu$, and with superconducting matter treated as a charged condensate coupled minimally to a classical electromagnetic four-potential $A_\mu$. Staticity implies hypersurface orthogonality of $\xi^\mu$ and allows a foliation by constant-time slices $\Sigma_t$. We denote the lapse by $\alpha=\sqrt{-\xi^2}$ and write the Schwarzschild exterior metric as in Section \ref{sec1}, with induced spatial metric $\gamma_{ij}$ on $\Sigma_t$ and future-pointing unit normal $n^\mu=\xi^\mu/\alpha$.

The condensate phase $\theta$ is not itself gauge invariant, but the combination \cite{London1935,tinkham1975introduction}
\begin{equation}
p_\mu \equiv \hbar \partial_\mu \theta - q A_\mu
\label{2.1}
\end{equation}
is well defined. For superconductors, we set $q=2e>0$ henceforth. The flux quantum is $\Phi_0=h/q$. The superflow four-velocity $u^\mu$ satisfies $u^\mu u_\mu=-1$ and the local Josephson relation identifies the projection of $p_\mu$ along $u^\mu$ with the chemical potential per particle \cite{Son:2000ht},
\begin{equation}
u^\mu p_\mu = \mu .
\label{2.2}
\end{equation}
Equations \eqref{2.1}-\eqref{2.2} are the kinematic core of our treatment. They encode both phase evolution and gauge coupling in a form that is independent of coordinates and gauge choices. All Josephson statements will be traced back to these invariants and to the projections determined by $\xi^\mu$ and $\Sigma_t$.

It is convenient to relate proper-time evolution in the local rest frame to measurements made by static observers. The latter have four-velocity $U^\mu=n^\mu=\xi^\mu/\alpha$, proper time $d\tau_{\rm stat}=\alpha dt$, and scalar electric potential $\Phi \equiv -U^\mu A_\mu$. When the banks are at rest with respect to the static frame, we may take $u^\mu=U^\mu$ within each bank up to small superflow corrections that enter only at higher order in gradients. In that quasi-static limit, the invariant Josephson relation becomes \cite{Son:2000ht,Gourgoulhon:2012ffd}
\begin{equation}
U^\mu \partial_\mu \theta = \frac{1}{\hbar}\left(\mu - q \Phi\right),
\qquad
\frac{d\theta}{dt} = \frac{\alpha}{\hbar}\left(\mu - q \Phi\right) .
\label{2.3}
\end{equation}
Equation \eqref{2.3} is the proper-to-asymptotic dictionary for phase evolution in a static geometry. Its content is that the Killing-time rate of the condensate phase is the redshifted electrochemical potential measured in the local rest frame. Our use of the lapse-normal split and hypersurface fluxes follows standard \(3{+}1\) treatments \cite{Gourgoulhon:2012ffd,Thorne_1982,Poisson:2009pwt}. The equilibrium redshift of intensive variables (electrochemical potential/temperature) is the Tolman-Ehrenfest law in GR thermodynamics \cite{Tolman:1930zza,Lima:2019brf,Wald:1999vt}. For the condensate, the covariant Josephson relation is consistent with relativistic hydrodynamics of a broken \(U(1)\) \cite{Son:2000ht}.

\subsection{Phase evolution, gauge invariants, and Tolman-Ehrenfest relations} \label{ssec2.1}
We begin by fixing notation for the gauge-invariant phase difference across a weak link and by recording the static-equilibrium redshift laws that constrain the intensive variables.

Let $\mathcal{C}$ be any curve that threads the junction from bank 1 to bank 2 within a constant-time slice $\Sigma_t$. The gauge-invariant phase drop is \cite{tinkham1975introduction,Barone_1982}
\begin{equation}
\Delta\varphi \equiv \theta_2 - \theta_1 - \frac{q}{\hbar}\int_{\mathcal{C}} A_idx^i ,
\label{2.4}
\end{equation}
where indices are raised and lowered with $\gamma_{ij}$ on $\Sigma_t$. Equation \eqref{2.4} is independent of the choice of $\mathcal{C}$ provided the magnetic flux through any two such curves differs by an integer multiple of the flux quantum. In simply connected regions free of vortices this independence is automatic. When the weak link is embedded in a loop, \eqref{2.4} is supplemented by the usual flux quantization condition, which we impose in Section \ref{sec5}.

To relate $\Delta\varphi$ to physical control parameters, we project the gauge-invariant momentum $p_\mu$ along $U^\mu$ and along directions tangent to $\Sigma_t$. The temporal projection is the local Josephson relation in the static frame,
\begin{equation}
U^\mu p_\mu = \hbar U^\mu\partial_\mu\theta - q U^\mu A_\mu
= \mu ,
\label{2.5}
\end{equation}
which is equivalent to the first equality in \eqref{2.3}. The spatial projection gives the superflow three-momentum on $\Sigma_t$,
\begin{equation}
p_i = \hbar \partial_i \theta - q A_i ,
\label{2.6}
\end{equation}
whose line integral across the barrier is precisely $\hbar \Delta\varphi$ up to the flux term already accounted for in \eqref{2.4}.

In static thermal equilibrium, the Tolman-Ehrenfest (TE) relations constrain the redshift of intensive quantities. The redshift mapping we use traces back to Tolman’s equilibrium condition and its modern refinements in GR thermodynamics \cite{Tolman:1930zza,Lima:2019brf,Wald:1999vt}, and our kinematic split follows standard $3+1$ treatments of static spacetimes \cite{Gourgoulhon:2012ffd,Poisson:2009pwt}. For a neutral fluid, $T,\alpha=\text{const}$ and $\mu,\alpha=\text{const}$ along a connected equilibrium configuration. For a charged condensate the relevant intensive is the electrochemical potential $\tilde{\mu}\equiv \mu - q \Phi$ measured in the local rest frame of the static observers. The equilibrium condition becomes \cite{Tolman:1930zza,Tolman:1930ona,Lima:2019brf}
\begin{equation}
\alpha\tilde{\mu} = \alpha(\mu - q \Phi) = \text{const on each bank} .
\label{2.7}
\end{equation}
Equation \eqref{2.7} follows either from maximization of the total entropy at fixed conserved charges in a static spacetime, or directly from the constancy of the Killing energy per particle. It implies that, in the absence of transport, any radial variation of $\Phi$ across a single bank is exactly compensated by the redshift of $\mu$.

The Josephson relation is a statement about departures from that strict equilibrium induced by weak biases. We parametrize those biases by the proper electrochemical potential drops $V^{\rm proper}_i$ applied to the two banks, defined with respect to the local static frames. Combining \eqref{2.3} with the TE background, the phase evolution in bank $i$ with respect to Killing time is \cite{Josephson:1962zz,tinkham1975introduction}
\begin{equation}
\frac{d\theta_i}{dt}=\frac{q}{\hbar} \alpha_i V^{\rm proper}_i,
\qquad \alpha_i\equiv \alpha(r_i) .
\label{2.8}
\end{equation}
The time derivative of the gauge-invariant phase drop $\Delta\varphi$ across the weak link is the difference of the two rates in \eqref{2.8}. This is the redshifted AC Josephson law that we analyze in detail in Section \ref{sec3}. Here, we note only two immediate consequences. First, equal proper biases $V^{\rm proper}_1=V^{\rm proper}_2$ do not imply equal asymptotic phase-evolution rates when $\alpha_1\neq \alpha_2$; rather, the higher-lapse terminal advances faster by a factor of $\alpha_i$. Second, in the weak-field limit with $\alpha_i\simeq 1+\Phi_N(r_i)/c^2$ in terms of the Newtonian potential $\Phi_N$, \eqref{2.8} reproduces the leading gravitational correction to the flat-space Josephson frequency relation.

Finally, it is useful to record the purely spatial identity that complements \eqref{2.8}. Taking a time derivative of \eqref{2.4} and using Maxwell’s equations together with Faraday’s law on $\Sigma_t$, one obtains
\begin{equation}
\frac{d}{dt}\Delta\varphi
= \frac{q}{\hbar}\Bigl(\alpha_1 V^{\rm proper}_1-\alpha_2 V^{\rm proper}_2\Bigr)
+\frac{q}{\hbar} \frac{d}{dt}\left(\frac{\Phi_{\rm loop}}{\Phi_0}\right) ,
\label{2.9}
\end{equation}
where $\Phi_{\rm loop}$ is the magnetic flux threading any loop completion of $\mathcal{C}$ and $\Phi_0=h/q$. For a single isolated junction, the second term is absent. In a loop geometry, it encodes the standard flux-to-phase conversion, while the first term carries the correctly redshifted bias. Equation \eqref{2.9} summarizes how gauge invariance, Killing symmetry, and the TE background cooperate to determine the phase dynamics that will be used in the AC analysis and in the interferometric discussion.

\subsection{Conserved currents and hypersurface fluxes} \label{ssec2.2}
The transport statements used later rest on the covariant continuity equation
\begin{equation}
\nabla_\mu j^\mu = 0 ,
\label{2.10}
\end{equation}
together with the static foliation by constant-$t$ hypersurfaces $\Sigma_t$ orthogonal to the Killing field. We decompose the current with respect to the static observers $U^\mu=\xi^\mu/\alpha$ by introducing the projector $h^\mu{}_{\nu}=\delta^\mu{}_{\nu}+U^\mu U_\nu$. The proper charge density and the spatial current measured in the static frame are
\begin{equation}
\rho \equiv -j^\mu U_\mu ,
\qquad
\mathcal{J}^\mu \equiv h^\mu{}_{\nu}j^\nu ,
\qquad
U_\mu \mathcal{J}^\mu=0 .
\label{2.11}
\end{equation}
On a slice $\Sigma_t$ with induced metric $\gamma_{ij}$, zero shift, and the split $j^\mu=\rho n^\mu+J^\mu$ (where $n_\mu J^\mu=0$), \eqref{2.10} is equivalent to the balance law \cite{Poisson:2009pwt,Gourgoulhon:2012ffd}
\begin{equation}
\partial_t\left(\sqrt{\gamma}\rho\right)
+ D_i\left(\sqrt{\gamma} \alpha\mathcal{J}^i\right)=0,
\label{2.12}
\end{equation}
where $D_i$ is the Levi-Civita connection of $\gamma_{ij}$. In \eqref{2.12} we used $\mathcal{L}_\xi\sqrt{\gamma}=\sqrt{\gamma} D_i(\alpha\beta^i)=0$ and $n^\mu=\xi^\mu/\alpha$. The quantity $\mathcal{F}^i\equiv \alpha\mathcal{J}^i$ is the flux density per unit Killing time. Equation \eqref{2.12} is the precise statement that the loss of charge inside any compact domain of $\Sigma_t$ per unit $t$ equals the outward flux of $\mathcal{F}^i$ through its boundary.

For any smooth two-surface $S\subset\Sigma_t$ with outward unit normal $s_i$ and area element $dA=\sqrt{\sigma}d^2x$, the conserved flux across $S$ per unit Killing time is
\begin{equation}
I_\infty[S] \equiv \int_{S} \mathcal{F}^i s_i  dA
= \int_S \alpha\mathcal{J}^i s_i  dA .
\label{2.13}
\end{equation}
The same flux expressed per unit proper time of the static observers at $S$ is
\begin{equation}
I_{\rm proper}[S] \equiv \int_{S} \mathcal{J}^i s_i  dA .
\label{2.14}
\end{equation}
When $\alpha$ is approximately constant over the junction area, \eqref{2.13}-\eqref{2.13} give the kinematic conversion
\begin{equation}
I_\infty[S] \simeq \alpha I_{\rm proper}[S] .
\label{2.15}
\end{equation}
This single factor of $\alpha$ reflects the time dilation between the proper clocks that register the local flow and the Killing time used by an asymptotic observer. In subsequent sections, we will combine \eqref{2.15} with the electrochemical driving redshift from Subsection \ref{ssec2.1} [Eq. \eqref{2.8}, where $V_i^\infty=\alpha_i V_i^{\rm proper}$ and $\dot\theta_i=(q/\hbar) V_i^\infty$]. The product of these two effects implies that currents measured at infinity scale as $\alpha$ relative to proper currents, while powers scale as $P_\infty=\alpha^2 P_{\rm proper}$; in particular, no additional $\alpha$ enters the intrinsic local critical current, and any $\alpha^2$ enhancement pertains to power (or to frequency-voltage products), not to $I_c$ itself.

It is often convenient to state the same relations intrinsically. Introducing the area covector $d\Sigma_\mu$ of $S\subset\Sigma_t$ by $d\Sigma_\mu = s_\mu dA$ with $s_\mu U^\mu=0$, the flux per unit Killing time can be written as
\begin{equation}
I_\infty[S] = \int_S \alpha j^\mu d\Sigma_\mu .
\label{2.16}
\end{equation}
Current conservation implies that $I_\infty[S]$ is insensitive to smooth deformations of $S$ that leave its boundary fixed, provided no sources or sinks are crossed. This surface-independence will be used in Section \ref{sec4} to evaluate asymptotic currents through geometrically simple surfaces, even when the microscopic transport occurs within a thin weak link of complicated shape.

Two further remarks will be used repeatedly. First, when the junction is small on the scale set by the background curvature, the fields and the lapse may be regarded as constant across the barrier up to corrections of order $(L/R_c)^2$, where $L$ is the junction size and $R_c$ the local curvature radius; in that regime, the mapping \eqref{2.15} is accurate. Second, in a loop geometry with two weak links on the same $\Sigma_t$, the net asymptotic current is obtained by summing the fluxes \eqref{2.13} through cross-sections that cut each link once; the result is independent of which cross-sections are chosen, and the phase constraint from flux quantization will be imposed independently in Section \ref{sec5}.

\subsection{Proper versus asymptotic parameters} \label{ssec2.3}
The calculations in later sections use a compact dictionary that translates local (proper) quantities defined in the static frame into the observables associated with Killing time. We collect the needed relations here and note their domains of validity.

Static observers have four-velocity $U^\mu=\xi^\mu/\alpha$ and proper time $d\tau_{\rm stat}=\alpha dt$. A proper frequency $\omega_{\rm proper}$ measured by such observers is related to the asymptotic frequency $\omega_\infty$ defined with respect to $t$ by \cite{Wald:1999vt,Gourgoulhon:2012ffd}
\begin{equation}
\omega_\infty=\alpha\omega_{\rm proper}.
\label{2.17}
\end{equation}
Conversely, a monochromatic signal of frequency $\Omega_\infty$ emitted from infinity is received at radius $r$ with proper frequency $\Omega_{\rm proper}=\Omega_\infty/\alpha(r)$. These relations are used both for the intrinsic phase evolution of a biased junction and for externally applied drives.

The electrochemical potential in the static frame is $\tilde{\mu}=\mu-q\Phi$. Combining the local Josephson relation with $d\tau_{\rm stat}=\alpha dt$ gives the conversion between a proper electrochemical drop $V^{\rm proper}$ and the corresponding phase-evolution rate per unit Killing time \cite{Josephson:1962zz,tinkham1975introduction},
\begin{equation}
\frac{d\theta}{dt}=\frac{q}{\hbar} \alpha V^{\rm proper}.
\label{2.18}
\end{equation}
Equation \eqref{2.18} is the only input needed to obtain the redshifted AC law in Section \ref{sec3} after taking a difference between the two banks. It also fixes the placement of the lapse factor throughout the frequency-voltage relations.

Spatial tensors decompose naturally on $\Sigma_t$ with induced metric $\gamma_{ij}$. Let $dA$ be the area element on a two-surface $S\subset\Sigma_t$ with unit normal $s_i$. The flux mapping derived in Subsection \ref{ssec2.2} implies
\begin{equation}
I_\infty[S]=\int_S \alpha\mathcal{J}^i s_idA 
\qquad
I_{\rm proper}[S]=\int_S \mathcal{J}^i s_idA 
\label{2.19}
\end{equation}
and, when $\alpha$ varies negligibly across $S$, the approximate relation
\begin{equation}
I_\infty[S]\simeq \alpha I_{\rm proper}[S].
\label{2.20}
\end{equation}
This expresses the time dilation between local transport and asymptotic readout. When combined with \eqref{2.18}, it implies that currents measured at infinity scale as $\alpha$ relative to proper currents, voltages scale as $\alpha$ relative to proper voltages, and powers scale as $P_\infty=\alpha^2 P_{\rm proper}$.

Throughout this work, asymptotic quantities (frequencies, voltages, currents) are defined with respect to the Killing time \(t\), the natural time coordinate for an observer at infinity. Real experiments, however, are performed by local observers at some finite radius \(r_{\rm lab}\), using their proper time \(\tau = \alpha(r_{\rm lab})\, t\). The basic conversions are:
\begin{equation}
    \omega_\infty = \alpha \,\omega_{\rm proper}, \quad V_\infty = \alpha \, V_{\rm proper}, \quad I_\infty = \alpha \, I_{\rm proper}.
\end{equation}
Here \(\omega_{\rm proper}\), \(V_{\rm proper}\), and \(I_{\rm proper}\) are measured using a laboratory clock and instruments at finite radius, whereas \(\omega_\infty\), \(V_\infty\), and \(I_\infty\) are the corresponding quantities expressed per unit Killing time. The AC Josephson relation uses whichever clock is chosen for phase comparison; the DC current uses whichever time parameter defines the flux. In weak-field terrestrial experiments, one always works with proper quantities, and the asymptotic ones differ only at \(O(g h / c^2)\).

For fields, it is convenient to distinguish invariant statements from gauge choices. The scalar potential seen by static observers is $\Phi=-U^\mu A_\mu$, while the electric and magnetic fields on $\Sigma_t$ are defined by $E_i=F_{i\mu}U^\mu$ and $B^i=\tfrac{1}{2}\epsilon^{ijk}F_{jk}$ with $\epsilon^{ijk}$ the Levi-Civita tensor of $\gamma_{ij}$. These definitions are gauge invariant and adapted to the foliation. The magnetic flux through a loop is $\Phi_{\rm loop}=\int_S B^i s_i dA$, independent of $\alpha$. Flux quantization therefore, retains its flat-space form on each static slice; redshift enters only through the time evolution of the junction phases, not through the spatial holonomy.

Length and area elements used in local constitutive relations are proper geometric quantities on $\Sigma_t$. If a weak link has proper thickness $L$ and proper cross-sectional area $A$, then the short-junction regime and any analytic estimates that rely on it are controlled by $L$ and by curvature radii computed from $\gamma_{ij}$. The asymptotic readout does not alter these geometric inputs. What it does alter is the conversion from proper current density to observed current flux via \eqref{2.19}-\eqref{2.20}, and from proper driving frequency to observed frequency via \eqref{2.17}.

Finally, when an RF drive of asymptotic frequency $\Omega_\infty$ and amplitude $\mathcal{V}_\infty(t)$ is imposed from infinity, the locally seen drive at radius $r$ is $\mathcal{V}_{\rm proper}(t_{\rm proper})=\mathcal{V}_\infty(t)/\alpha(r)$ with $t_{\rm proper}=\alpha(r)t$, up to propagation effects that depend on the particular geometry of the field lines and are treated separately in the appendices. Substituting this into \eqref{2.18} shows that, expressed in Killing time, the phase locking has the flat-space form with the replacement $V\mapsto \mathcal{V}_\infty$ and drive frequency $\Omega\mapsto \Omega_\infty$; if one instead works entirely in the local proper frame, then the corresponding replacements are $V\mapsto \mathcal{V}_{\rm proper}$ and $\Omega\mapsto \Omega_\infty/\alpha$. This closes the dictionary between proper and asymptotic parameters used in Sections \ref{sec3}-\ref{sec5}.

\section{AC Josephson effect with gravitational redshift in Schwarzschild} \label{sec3}
We now analyze phase dynamics across a weak link whose superconducting banks are static at radii \(r_1\) and \(r_2\) in the Schwarzschild exterior. The junction is assumed short on the curvature scale so that fields and lapse vary negligibly over its area. The control parameters are small electrochemical biases applied to each bank in their local static frames, and there is no external time-dependent drive in this subsection. The objective is to obtain a gauge-invariant relation between the time derivative of the phase drop, defined with respect to the Killing time \(t\), and the proper biases on the two banks. The essential ingredients have already been established: the gauge-invariant phase difference on a static slice and the redshift conversion between proper-time evolution and asymptotic rates.

\subsection{Redshifted AC law} \label{ssec3.1}
Let \(\Delta\varphi\) be the gauge-invariant phase drop across the weak link defined on \(\Sigma_t\) as in \eqref{2.4}. In the absence of any explicitly time-dependent vector potential in the junction region, Faraday’s law on \(\Sigma_t\) implies that the magnetic-flux contribution to \(\dot{\Delta\varphi}\) vanishes. Using the local relation \eqref{2.18} for each bank, the Killing-time derivatives of the phases in the two electrodes are
\begin{equation}
\frac{d\theta_i}{dt}=\frac{q}{\hbar} \alpha_i V_i^{\mathrm{proper}},
\qquad
\alpha_i\equiv \alpha(r_i),\ \ i=1,2 .
\label{3.1}
\end{equation}
Taking the difference of the two rates gives the redshifted AC Josephson law
\begin{equation}
\dot{\Delta\varphi}_\infty
= \frac{2e}{\hbar}\left(\alpha_1 V_1^{\mathrm{proper}}
- \alpha_2 V_2^{\mathrm{proper}}\right).
\label{3.2}
\end{equation}
The frequency \(\dot{\Delta\varphi}_\infty\) in Eq. \eqref{3.2} is the phase-evolution rate with respect to Killing time. A local laboratory at finite radius measures instead the proper frequency \(\dot{\Delta\varphi}_{\rm proper} = \dot{\Delta\varphi}_\infty / \alpha(r_{\rm lab})\). When comparing the Josephson oscillation to a local reference clock, one must use \(\dot{\Delta\phi}_{\rm proper}\); when comparing to a reference defined at infinity (or a long-baseline frequency standard), \(\dot{\Delta\varphi}_\infty\) is appropriate. The two differ only by the redshift factor \(\alpha\).

\begin{figure}
    \centering
    \includegraphics[width=\columnwidth]{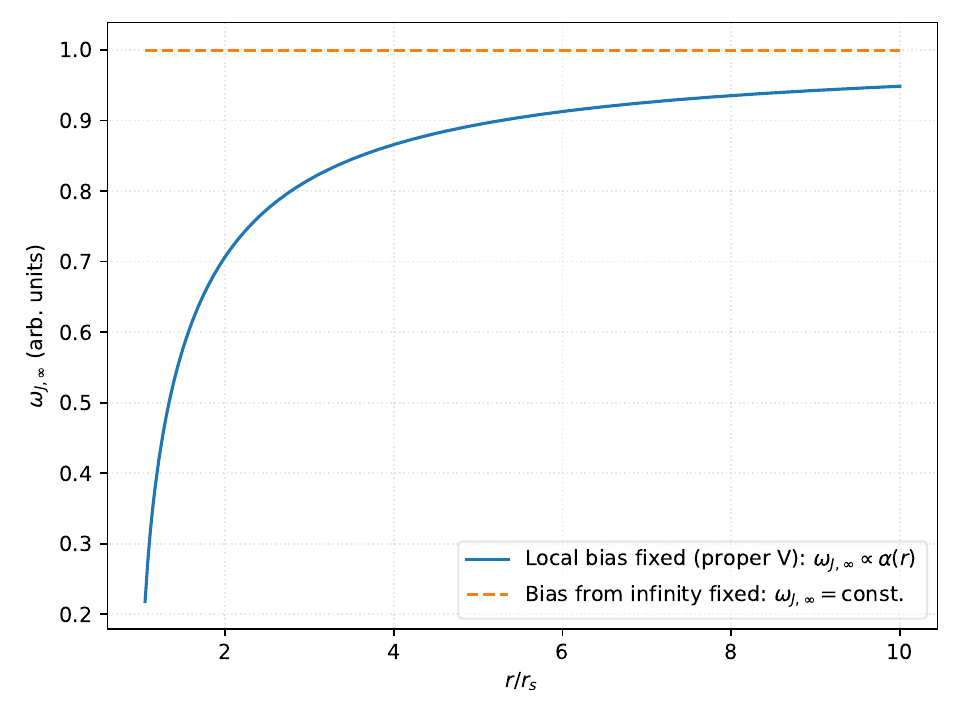}
    \caption{AC redshift map. Josephson frequency observed at infinity, \(\omega_{J\infty}\), versus radius \(r/r_s\) under two bias conventions: with a fixed local (proper) bias, \(\omega_{J\infty}\propto \alpha(r)=\sqrt{1-r_s/r}\) and is redshift-suppressed near the horizon; with a fixed asymptotic bias, \(\omega_{J\infty}\) is independent of \(\alpha\) and remains constant.}
    \label{fig1}
\end{figure}

Figure \ref{fig1} isolates the kinematic role of the lapse in the AC effect. If the same numerical voltage is applied locally in proper time, the Josephson relation \( \hbar d\Delta\varphi/d\tau = 2eV_{\rm proper} \) and \(d\tau=\alpha dt\) imply \( \hbar d\Delta\varphi/dt = 2e\alpha V_{\rm proper} \), so the frequency measured at infinity scales like \(\alpha(r)\) and vanishes at the horizon. By contrast, when both the dc bias and the rf reference are specified at infinity, the local drop is \(V_{\rm proper}=V_\infty/\alpha\) and the observed frequency is \( \hbar d\Delta\varphi/dt=2eV_\infty\), independent of position. Figure \ref{fig1} therefore, makes two points clear: redshift is an operational statement about how the bias is defined, and metrological comparisons performed in asymptotic variables recover the flat-space AC law exactly.

Note that the Josephson relation is the time derivative of the
gauge-invariant phase difference, i.e. $\dot{\Delta\varphi}_\infty$ can be read as “difference of the individual time derivatives”; it is only implicitly equivalent when both phases are evaluated at the same Killing time. The form $\dot{\Delta\varphi}$ is standard and unambiguous. Equation \eqref{3.2} is manifestly gauge invariant, since it is written entirely in terms of the electrochemical drops measured in the local static frames and the lapse factors determined by the timelike Killing field. It reduces to the flat-space relation when \(\alpha_1=\alpha_2=1\). In the flat-space limit, this reduces to the standard Josephson frequency-voltage law and its microwave locking (Shapiro steps) \cite{Josephson:1962zz,Shapiro:1963nhj,Langenberg_1966,tinkham1975introduction}; numerical values use CODATA constants for \(2e/h\) \cite{Tiesinga_2021}.

Two biasing conventions are worth separating because they lead to distinct scalings near the horizon. If the biases are imposed locally by agents comoving with each bank, then \(V_i^{\mathrm{proper}}\) are the primary control parameters and \eqref{3.2} displays the explicit suppression of the asymptotic frequency as \(r_i\to r_s\), since \(\alpha_i\to 0\). If instead the biases are delivered from infinity along static leads, then the proper drops scale as \(V_i^{\mathrm{proper}}=V_i^{\infty}/\alpha_i\). Substituting into \eqref{3.2} yields
\begin{equation}
\dot{\Delta\varphi}_\infty
= \frac{2e}{\hbar}\left(V_1^{\infty}-V_2^{\infty}\right),
\label{3.3}
\end{equation}
which is independent of redshift. Both conventions are physically realizable and will be kept in view when we discuss limiting regimes. Writing the weak-field lapse as \(\alpha_i\simeq 1+\Phi_N(r_i)/c^2\) with the Newtonian potential \(\Phi_N = -GM/(r_ic^2)\) and \(|\Phi_N|\ll c^2\), one finds
\begin{align}
\dot{\Delta\varphi}_\infty
&\simeq \frac{2e}{\hbar}\Biggl[
V_1^{\mathrm{proper}}-V_2^{\mathrm{proper}} \nonumber \\
&+\frac{V_1^{\mathrm{proper}}\Phi_N(r_1)-V_2^{\mathrm{proper}}\Phi_N(r_2)}{c^2} \Biggr] .
\label{3.4}
\end{align}
For equal proper biases \(V_1^{\mathrm{proper}}=V_2^{\mathrm{proper}}=V\) this becomes
\begin{equation}
\dot{\Delta\varphi}_\infty
\simeq \frac{2e}{\hbar} \frac{V}{c^2}\left[\Phi_N(r_1)-\Phi_N(r_2)\right],
\label{3.5}
\end{equation}
which is the leading gravitational shift of the Josephson frequency. In a uniform field over a vertical separation \(\Delta h\), one may replace \(\Phi_N(r_1)-\Phi_N(r_2)\) by \(g \Delta h\). The gravitationally induced frequency shift predicted here is consistent with weak-field Josephson analyses \cite{Ummarino:2020loo} and complementary to precision redshift tests using atomic clocks and atom interferometers \cite{Pumpo:2021sok,Zheng:2022hwj}.

Near the horizon, the Rindler approximation gives \(\alpha(r)\simeq \kappa \rho\), where \(\kappa\) is the surface gravity and \(\rho\) the proper distance to the horizon along a static ray \cite{Wald:1999vt}. For fixed local biases, this produces
\begin{equation}
\dot{\Delta\varphi}_\infty
\simeq \frac{2e}{\hbar}\left(\kappa\rho_1 V_1^{\mathrm{proper}}
-\kappa\rho_2 V_2^{\mathrm{proper}}\right),
\label{3.6}
\end{equation}
exhibiting linear suppression as either bank approaches the horizon \((\rho\to 0)\). Under the “bias-from-infinity” convention, \eqref{3.3} shows that the asymptotic frequency remains finite in the same limit.

Finally, we note two consistency checks. First, if the banks are at the same radius \(r_1=r_2\) and are biased identically in their local frames, then \(\dot{\Delta\varphi}_\infty=0\) and there is no AC signal at infinity, in agreement with intuition. Second, any static gauge transformation \(A_\mu\to A_\mu+\partial_\mu\chi\), \(\theta\to \theta+\tfrac{q}{\hbar}\chi\) leaves \eqref{3.2} unchanged because \(V_i^{\mathrm{proper}}\) are physical drops of electrochemical potential and the derivation uses only the invariant projections defined in Section \ref{sec2}. These properties will be important when we superimpose periodic driving from infinity in Subsection \ref{ssec3.2}, where the redshift of the drive frequency and propagation phases become essential.

\subsection{RF drive from infinity and step positions} \label{ssec3.2}
We now superimpose a weak monochromatic drive generated at infinity with angular frequency $\Omega_\infty$. The drive propagates through the Schwarzschild exterior and impinges on the two banks located at $r_1$ and $r_2$. Two effects must be accounted for: the gravitational redshift of the local proper frequency and amplitude seen by static observers at each bank, and the propagation phase accumulated between infinity and $r_i$ (including the leading Shapiro delay). The RF phase offset includes geometric path differences and the Shapiro time delay \cite{Shapiro:1964uw,Possel:2019zbu}; gravitational AB-type phases have been discussed and observed in matter-wave settings \cite{Overstreet:2021hea,Chiao:2023ezj}.
 The latter will enter only through a phase offset; closed expressions for that offset are recorded in the appendices. 

To allow for general coupling of the external field to the two banks, we write the proper electrochemical drops as
\begin{equation}
V_i^{\mathrm{proper}}(t)
= V_{i,\rm dc}^{\mathrm{proper}} + \frac{1}{\alpha_i}\mathrm{Re}\Big\{\kappa_iV_{\rm rf}^{\infty}e^{,i(\Omega_\infty t-\psi_i)}\Big\},
\label{3.7}
\end{equation}
where $i=1,2,..$, and $\kappa_i$ encode the (dimensionless) coupling of the asymptotic drive to bank $i$, $\psi_i$ are the propagation phases from infinity to $r_i$, and $\alpha_i=\alpha(r_i)$. The factor $1/\alpha_i$ converts an asymptotic voltage to the proper voltage seen by static observers at $r_i$ (since $V_i^{\infty}=\alpha_i V_i^{\rm proper}$).

Inserting \eqref{3.7} into the gauge-invariant phase-evolution law \eqref{3.2} and integrating once in time yields
\begin{equation}
\Delta\varphi(t)=\omega_0t + a\sin\left(\Omega_\infty t-\psi\right)+\varphi_0 ,
\label{3.8}
\end{equation}
with
\begin{equation}
\omega_0=\frac{2e}{\hbar}\left(\alpha_1 V_{1,\rm dc}^{\mathrm{proper}}
-\alpha_2 V_{2,\rm dc}^{\mathrm{proper}}\right),
\quad
a=\frac{2e}{\hbar} \frac{|V_{\rm rf,eff}^{\infty}|}{\Omega_\infty},
\label{3.9}
\end{equation}
\begin{equation}
V_{\rm rf,eff}^{\infty}\equiv \kappa_1 V_{\rm rf}^{\infty} e^{-i\psi_1}-\kappa_2 V_{\rm rf}^{\infty} e^{-i\psi_2},
\label{3.9a}
\end{equation}
and $\psi=\arg V_{\rm rf,eff}^{\infty}$. The effective drive experienced by the junction is the differential combination $V_{\rm rf,eff}^{\infty}$; when the drive is purely common-mode, $V_{\rm rf,eff}^{\infty}=0$ and no time-dependent phase is induced, as expected.

Assuming the short-junction current-phase relation $I(t)=I_c\sin\Delta\varphi(t)$, the time average $\langle I\rangle$ can be obtained by the Jacobi-Anger expansion. A nonzero DC component arises precisely when the intrinsic frequency $\omega_0$ locks to an integer multiple of the drive frequency:
\begin{equation}
\omega_0 = n \Omega_\infty,\qquad n\in \mathbb{Z}.
\label{3.10}
\end{equation}
This is the invariant step condition. Written directly in terms of control parameters,
\begin{equation}
\Big\langle \alpha_1 V_1^{\mathrm{proper}} - \alpha_2 V_2^{\mathrm{proper}} \Big\rangle
= n \frac{\hbar\Omega_\infty}{2e}.
\label{3.11}
\end{equation}
When the DC bias is delivered from infinity across the two banks (so that $V_{i,\rm dc}^{\mathrm{proper}}=V_{i,\rm dc}^{\infty}/\alpha_i$), \eqref{3.10} becomes
\begin{equation}
\frac{2e}{\hbar} (V_{1,\rm dc}^{\infty} - V_{2,\rm dc}^{\infty})=n \Omega_\infty,
\quad\Longrightarrow\quad
\langle V_\infty\rangle = n \frac{\hbar\Omega_\infty}{2e},
\label{3.12}
\end{equation}
which is the familiar flat-space Shapiro step locus expressed entirely in asymptotic observables and thus independent of redshift. In contrast, when the DC bias is imposed locally in the proper frames, the step locus retains the explicit $\alpha_i$ factors through $\omega_0$ as in \eqref{3.9}.

The step heights depend on the effective modulation index $a$ and on the propagation phases only through the complex amplitude $V_{\rm rf,eff}^{\infty}$. Carrying out the averaging with $\Delta\varphi(t)$ from \eqref{3.8} gives
\begin{equation}
\langle I\rangle = I_cJ_n(a)\sin(\varphi_0-n\psi),
\label{3.13}
\end{equation}
on the $n$-th step, where $J_n$ is the Bessel function of the first kind \cite{Shapiro:1963nhj,Langenberg_1966,tinkham1975introduction}. The magnitude $|J_n(a)|$ controls the lobe envelope, while the phase $\psi$ encodes the net propagation phase difference between the two banks. To leading post-Newtonian order, $\psi_1-\psi_2=\Omega_\infty\Delta t_{\rm prop}$, with $\Delta t_{\rm prop}$ the differential coordinate-time delay; explicit expressions in terms of the Schwarzschild radius and the geometric path parameters are relegated to the appendices.

\begin{figure}
    \centering
    \includegraphics[width=\columnwidth]{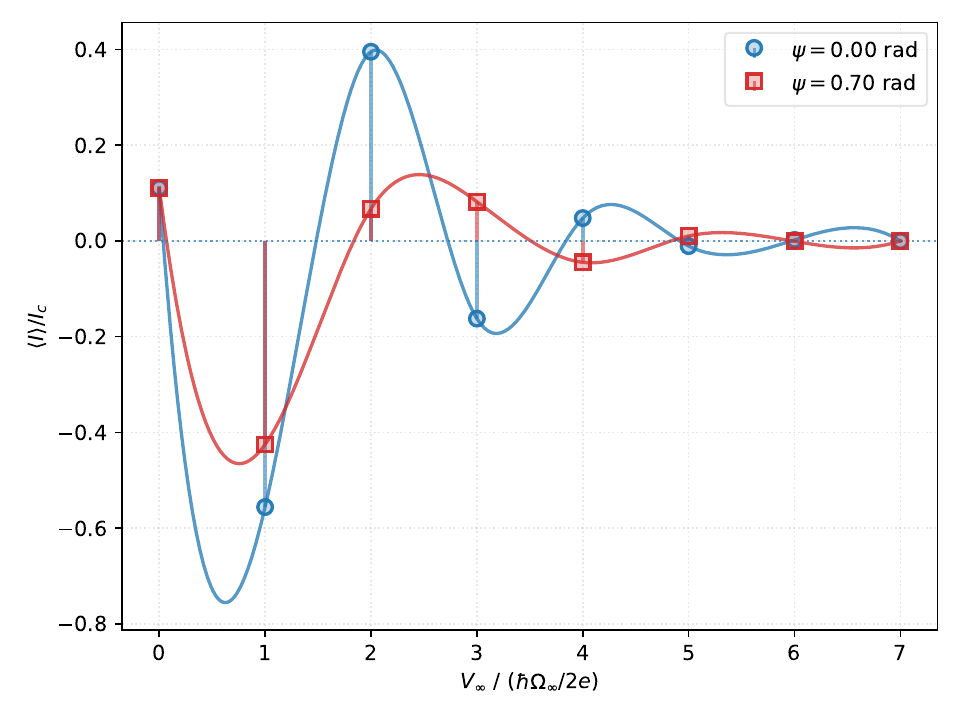}
    \caption{Shapiro steps in asymptotic variables. Time-averaged Josephson current \(\langle I\rangle/I_c\) versus asymptotic dc bias \(V_\infty/(\hbar\Omega_\infty/2e)\). With both bias and drive defined at infinity, the step loci sit at integers \(V_\infty = n \hbar\Omega_\infty/2e\) (vertical positions on the x-axis), independent of gravitational potential. The step heights are modulated by the propagation phase \(\psi\) through Bessel coefficients \(J_n(a)\) and the factor \(\cos(n\psi)\); two choices of \(\psi\) illustrate that only heights change, not positions.}
    \label{fig2}
\end{figure}

In Figure \ref{fig2}, the asymptotic-variable formulation of the AC effect is shown. Writing the phase as \(\Delta\varphi(t)=\Delta\varphi_0 + (2e/\hbar)V_\infty t + a\sin(\Omega_\infty t+\psi)\), the time average \(\langle I\rangle = I_c\langle\sin\Delta\varphi\rangle\) vanishes except when the Josephson frequency commensurates with the drive, \((2e/\hbar)V_\infty = n\Omega_\infty\). At these discrete loci, the average picks out a single Fourier component, yielding \(\langle I\rangle \propto J_n(a)\sin(\Delta\varphi_0 - n\psi)\). Fixing \(\Delta\varphi_0=\pi/2\) highlights the \(\cos(n\psi)\) modulation of the step heights. The essential outcome is metrological: once the dc bias and rf reference are specified at infinity, the locations of the steps in \(V_\infty\) coincide with their flat-space values, while gravitational influence appears only through the propagation phase \(\psi\) that shapes the envelope.

Two limiting regimes are instructive. In the weak-field limit, $\alpha_i\simeq 1+\Phi_N(r_i)/c^2$ (with Newtonian potential $\Phi_N$ and $|\Phi_N|\ll c^2$) and the differential propagation phase reduces to the flat-space geometric phase plus a small, logarithmic Shapiro correction. The step positions remain at $\langle V_\infty\rangle=n\hbar\Omega_\infty/2e$ for asymptotically delivered bias, while the lobe heights acquire only $\mathcal{O}(\Phi_N/c^2)$ corrections through $a$ and $\psi$. In the near-horizon (Rindler) regime, the local proper frequency of the incoming drive is $\Omega_{\rm proper}(r)=\Omega_\infty/\alpha(r)$ and thus becomes large, but the step locus in terms of asymptotic DC voltage continues to be given by \eqref{3.12} for drive and bias defined at infinity; all near-horizon enhancements reside in the modulation index $a\propto |V_{\rm rf,eff}^{\infty}|/\Omega_\infty$ and in the differential phase through $\psi$.

Equations \eqref{3.10}-\eqref{3.13} summarize the RF-driven AC Josephson phenomenology in Schwarzschild. They display how gravitational redshift and propagation affect the phase dynamics only through the proper-asymptotic conversion and the propagation phases, while leaving the asymptotic step loci invariant when both the DC bias and the RF drive are specified at infinity.

\subsection{Limiting regimes} \label{ssec3.3}
The formulas derived above admit transparent expansions in two opposite regimes that will be used repeatedly later: a weak-field domain in which the lapse $\alpha$ varies mildly across the device, and a near-horizon (Rindler) domain in which $\alpha$ is small and varies rapidly with radius. We also record the degenerate case $r_1=r_2$, which provides a useful check.

In the weak-field limit \(|\Phi_N|/c^2\ll 1\), the expansion of \eqref{3.2} is already given in \eqref{3.4}, including the equal-bias specialization; see also the step conditions \eqref{3.10}-\eqref{3.13}. Tidal corrections to any of these statements scale as $(L/R_c)^2$, with $L$ the junction size and $R_c$ the local curvature radius determined by the Schwarzschild mass parameter. Throughout, we assume $L\ll R_c$, so these effects can be neglected at leading order.

Next, for the near-horizon approximation (Rindler), let $r_s=2GM/c^2$ and $\delta r\ll r_s$. The lapse behaves as $\alpha(r)\simeq \kappa \rho$, where $\rho$ is the proper distance from the horizon along a static ray and $\kappa=c^4/(4GM)$ is the surface gravity \cite{Wald:1999vt}. With locally imposed proper biases, \eqref{3.2} becomes
\begin{equation}
\dot{\Delta\varphi}_\infty
\simeq \frac{2e}{\hbar}\left(\kappa\rho_1 V_1^{\rm proper}-\kappa\rho_2 V_2^{\rm proper}\right),
\label{3.17}
\end{equation}
exhibiting linear suppression as either bank approaches the horizon ($\rho\to 0$). In contrast, when both DC bias and RF drive are specified at infinity, \eqref{3.12} shows that step positions remain finite and independent of $\alpha$; the near-horizon behavior is then confined to the envelope via the large local drive frequency $\Omega_{\rm proper}(r)=\Omega_\infty/\alpha(r)$ and to the phase offset $\psi$, which acquires the leading Rindler propagation phase.

For the DC mapping, see Sec. \ref{sec4} and especially \eqref{4.16}-\eqref{4.17}. Finally, we check the equal-radii case and bias symmetry. If $r_1=r_2$, then $\alpha_1=\alpha_2$ and the redshift factors drop out of \eqref{3.2}. For equal proper biases, the AC signal vanishes, while for a differential proper bias (or, equivalently, for a differential asymptotic bias), the frequency is exactly the flat-space value. Under RF drive, common-mode coupling $\kappa_1=\kappa_2$ eliminates the time-dependent phase in \eqref{3.9}, consistent with the interpretation that only differential driving reaches the junction. These checks ensure that redshift enters solely through proper-asymptotic conversions and propagation phases, in agreement with the gauge-invariant construction of Section \ref{sec2}.

\section{DC weak-link transport and the redshift mapping of \texorpdfstring{$J_{c}$}{Jc}} \label{sec4}
We now turn to stationary transport across a short S-N-S weak link embedded in a Schwarzschild exterior. The background is static, the junction is at rest with respect to the static observers, and all quantities are time independent in the local frame. Our strategy is two-step. First, we formulate the superconducting condensate on a constant-$t$ hypersurface $\Sigma_t$ using a covariant Ginzburg-Landau (GL) functional \cite{Ginzburg:1950sr} and derive the local current-phase relation in proper geometric variables. This furnishes a proper critical current $I_c^{\rm proper}$ (and current density $J_c^{\rm proper}$) expressed solely in terms of intrinsic junction parameters: proper thickness, proper cross-sectional area, coherence lengths, and interface transparencies. Second, in Subsection \ref{ssec4.2} we map that local result to the asymptotic flux by the hypersurface construction developed in Section \ref{sec2}, thereby obtaining the one-power redshift scaling $I_{c,\infty}\simeq \alpha I_c^{\rm proper}$ (and $J_{c,\infty}\simeq \alpha J_c^{\rm proper}$ when the area is the same). This cleanly separates local constitutive physics from kinematic redshift and time dilation. We emphasize that the $\alpha^2$ factor applies to power ($P_\infty=\alpha^2 P_{\rm proper}$), not to current.

\subsection{Local GL formulation on a static slice} \label{ssec4.1}
Let $\Sigma_t$ be a static slice with induced metric $\gamma_{ij}$, Levi-Civita connection $D_i$, and area/volume elements built from $\sqrt{\gamma}$. The superconducting condensate is described by $\psi=|\psi|e^{i\theta}$ minimally coupled to the spatial vector potential $A_i$ pulled back to $\Sigma_t$. Throughout this subsection, the electromagnetic sector serves only as a classical background; we ignore its backreaction on the metric. To avoid confusion with the lapse, we denote the quadratic GL coefficient by $a(T)$ and the quartic by $b>0$. The local constitutive input is the London-GL framework \cite{London1935,tinkham1975introduction}; see also recent discussions of superconductors in static gravitational/electromagnetic backgrounds \cite{Ummarino:2021tpz,Ummarino:2021vwc} and a curved-spacetime GL proposal \cite{Rastkhadiv:2025ibd}.

A covariant GL free-energy functional on $\Sigma_t$ is
\begin{align}
\mathcal{F}[\psi,A_i]
&=\int_{\Sigma_t} d^3x \sqrt{\gamma} 
\Biggl[
a|\psi|^{2}+\frac{b}{2}|\psi|^{4} \nonumber \\
&+\frac{\hbar^{2}}{2m_*}\gamma^{ij}(D_i\psi)^*(D_j\psi)
\Biggr],
\label{4.1}
\end{align}
with covariant gauge derivative
\begin{equation}
D_i\psi=\left(D_i-\frac{i q}{\hbar}A_i\right)\psi .
\label{4.2}
\end{equation}
Varying \eqref{4.1} with respect to $\psi^*$ gives the GL equation on $\Sigma_t$,
\begin{equation}
-\frac{\hbar^{2}}{2m_*}\frac{1}{\sqrt{\gamma}}D_i\left(\sqrt{\gamma} \gamma^{ij}D_j\psi\right)
+a\psi+b|\psi|^2\psi=0.
\label{4.3}
\end{equation}

The gauge-invariant spatial current density measured by the static observers is obtained by varying $\mathcal{F}$ with respect to $A_i$,
\begin{equation}
\mathcal{J}^{i}
=\frac{q\hbar}{m_*}\text{Im}\left(\psi^{*}\gamma^{ij}D_j\psi\right)
D_i \mathcal{J}^{i}=0
\label{4.4}
\end{equation}
in the stationary regime. Equations \eqref{4.3}-\eqref{4.4} are intrinsic to $\Sigma_t$ and require no coordinate choice beyond the static foliation. 

We model the weak link as a thin normal (or weakly superconducting) barrier occupying a slab $0<\ell<L$ bounded by superconducting banks at $\ell<0$ and $\ell>L$. Here $\ell$ is the proper distance across the barrier along a geodesic orthogonal to the junction interfaces, and $A$ denotes the proper cross-sectional area. Inside the barrier, pair breaking makes $|\psi|$ small, so the GL equation can be linearized with a positive quadratic coefficient $a_N>0$ \cite{Gennes_1966}. Taking $A_i$ slowly varying across the thickness and choosing a gauge where $A_\ell$ is constant in the barrier, the linearized equation reduces to
\begin{equation}
-\xi_N^{2}\partial_{\ell}\left(\sqrt{\gamma} \gamma^{\ell\ell}\partial_{\ell}\psi\right)\big/\sqrt{\gamma}
+\psi=0,
\quad
\xi_N^{2}\equiv \frac{\hbar^{2}}{2m_* a_N}.
\label{4.5}
\end{equation}
When the junction is short on curvature scales, $\gamma^{\ell\ell}$ and $\sqrt{\gamma}$ may be treated as constants across the barrier up to corrections of order $(L/R_c)^2$ \cite{Barone_1982}. Equation \eqref{4.5} then yields the familiar exponential profile,
\begin{equation}
\psi(\ell)=\psi_L\frac{\sinh\left[(L-\ell)/\xi_N\right]}{\sinh(L/\xi_N)}
+\psi_R\frac{\sinh\left(\ell/\xi_N\right)}{\sinh(L/\xi_N)},
\label{4.6}
\end{equation}
where $\psi_L$ and $\psi_R$ are the order parameters at the two interfaces, to be related to the bank amplitudes by interface boundary conditions. Writing $\psi_{L R}=|\psi_{L R}|e^{i\theta_{L R}}$ and inserting \eqref{4.6} into \eqref{4.4}, the uniform current density through the slab is (to leading order in $|\psi|$)
\begin{equation}
\mathcal{J}^{\ell}
=\frac{q\hbar}{m_*}\gamma^{\ell\ell}
\frac{|\psi_L||\psi_R|}{\xi_N\sinh(L/\xi_N)}
\sin\left(\theta_R-\theta_L-\frac{q}{\hbar}\int_0^LA_\ell d\ell\right).
\label{4.7}
\end{equation}
The quantity in parentheses is precisely the gauge-invariant phase drop across the barrier defined on $\Sigma_t$. Denoting it by $\Delta\varphi$, the current through the junction is therefore
\begin{equation}
I_{\rm proper}= \int_{S}\mathcal{J}^{\ell}dA
= I_c^{\rm proper}\sin\Delta\varphi ,
\label{4.8}
\end{equation}
with the proper critical current
\begin{equation}
I_c^{\rm proper}
=\frac{q\hbar}{m_*}\gamma^{\ell\ell}
\frac{|\psi_L||\psi_R|}{\xi_N\sinh(L/\xi_N)}A .
\label{4.9}
\end{equation}
Equations \eqref{4.8}-\eqref{4.9} are the local current-phase relation and critical current expressed solely in proper geometric data on $\Sigma_t$. Two comments clarify their domain of validity. First, short-junction conditions require $L\ll \xi_{L},\xi_{R}$ in the banks and negligible variation of $\gamma_{ij}$ across the barrier. Second, interface physics is encoded in the effective values of $|\psi_{L R}|$ that appear in \eqref{4.9}. One may adopt de Gennes-type boundary conditions to relate $|\psi_{L R}|$ to the bulk bank amplitudes and interface transparencies, but the subsequent mapping to asymptotic observables is insensitive to those microscopic details.

It is useful to isolate the leading thickness dependence. For $L\gtrsim \xi_N$, the factor $1/\sinh(L/\xi_N)$ produces the standard exponential suppression $I_c^{\rm proper}\propto e^{-L/\xi_N}$. For $L\ll \xi_N$, one has $I_c^{\rm proper}\propto A \gamma^{\ell\ell}|\psi_L||\psi_R|/L$. In either regime, the current-phase relation remains sinusoidal at leading order, a fact we will use in Section \ref{sec5} when analyzing interferometry. The passage from $I_c^{\rm proper}$ to the asymptotically measured $I_{c,\infty}$ (or $J_{c,\infty}$) proceeds via the hypersurface flux mapping and the redshift of electrochemical driving derived in Section \ref{sec2}. Combining those ingredients (specifically, $I_\infty\simeq \alpha I_{\rm proper}$ when $\alpha$ is uniform across the section, and $V^\infty=\alpha V^{\rm proper}$) yields the one-power $\alpha$ law in Subsection \ref{ssec4.2}. In the weak-curvature, short-junction regime ($L\ll R_c$), spatial metric factors vary only by $\mathcal{O}((L/R_c)^2)$ across the barrier; treating $\sqrt{\gamma} \gamma^{\ell\ell}$ as constant through the barrier therefore induces only $\mathcal{O}((L/R_c)^2)$ corrections to $I_c$.

\subsection{From local solution to asymptotic current} \label{ssec4.2}
The goal is to convert the proper current-phase relation obtained on $\Sigma_t$ into the current that an observer at infinity would infer from a stationary DC experiment. The only ingredients are the hypersurface flux mapping of Section \ref{sec2} and the redshift of electrochemical driving.

Let $S\subset\Sigma_t$ be any smooth cross-section that intersects the weak link once and is orthogonal to the transport direction $\ell$. In the stationary regime, the conserved flux of charge through $S$ per unit Killing time is
\begin{equation}
I_\infty[S]=\int_S \alpha\mathcal{J}^{i}s_i dA ,
\label{4.10}
\end{equation}
with $s_i$ the unit normal on $\Sigma_t$ and $dA$ the proper area element. When $\alpha$ varies negligibly across the junction area, this reduces to the kinematic conversion
\begin{equation}
I_\infty \simeq \alpha I_{\rm proper},
\qquad
I_{\rm proper}=\int_S \mathcal{J}^{\ell}dA .
\label{4.11}
\end{equation} 
We stress the distinction between proper and asymptotic currents. The conserved flux measured by a laboratory at finite radius is \(I_{\rm proper}\), the charge crossing a surface per unit proper time. The corresponding quantity per unit Killing time is \(I_\infty = \alpha I_{\rm proper}\). The asymptotic critical current \(I_{c,\infty}\) used here is an invariant characterization of the junction and is related to the proper, locally measured critical current by a single factor of the lapse.

The one-lapse scaling in Eq. \eqref{4.11} is independent of junction microphysics and complements gravity-based Josephson implementations on the holographic side \cite{Horowitz:2011dz} and broader reviews on superconductors in gravity \cite{Gallerati:2022pgh}. Substituting the local current-phase relation \eqref{4.8} gives, to leading order in junction shortness and curvature
\begin{equation}
I_\infty \simeq \alpha I_c^{\rm proper}\sin\Delta\varphi .
\label{4.12}
\end{equation}
Two immediate consequences follow. First, the asymptotic DC critical current, defined as the maximum charge flux per unit Killing time at fixed junction phase, reads
\begin{equation}
I_{c,\infty}^{\rm (phase)} \simeq \alpha I_c^{\rm proper}
\quad \text{(fixed $\Delta\varphi$).}
\label{4.13}
\end{equation}
Second, the conversion is independent of microscopic details of the barrier, which have already been absorbed into $I_c^{\rm proper}$ through \eqref{4.9}; curvature corrections are $\mathcal{O}((L/R_c)^2)$.

In many DC protocols, however, the junction is not phase-clamped locally. Instead, one controls the junction by an electrochemical bias delivered from infinity and uses the Josephson relation to hold the phase static. To make contact with that operational definition, we combine \eqref{4.12} with the redshift of the bias-to-phase conversion. Stationarity of the phase requires the vanishing of the Killing-time phase rate,
\begin{align}
\dot{\Delta\varphi}_\infty
&=\frac{2e}{\hbar} \left(\alpha_1 V_1^{\rm proper}-\alpha_2 V_2^{\rm proper}\right) \nonumber \\
&=\frac{2e}{\hbar} \left(V_1^{\infty}-V_2^{\infty}\right)=0 .
\label{4.14}
\end{align}
If the DC bias is delivered from infinity along static leads, then $V_i^{\rm proper}=V_i^{\infty}/\alpha_i$ and \eqref{4.14} enforces $V_1^{\infty}=V_2^{\infty}$, as expected for a dissipationless stationary state. Small departures from stationarity provide a controlled way to infer the critical current: one perturbs the electrochemical balance by a slowly ramped asymptotic bias $\delta V_\infty$, which produces a small phase-slip rate via the redshifted Josephson relation. Linearizing in that bias, using $\frac{d\Delta\varphi}{dt}=\frac{2e}{\hbar} \delta V_\infty$ (equivalently $\frac{d\Delta\varphi}{dt}=\frac{2e}{\hbar} \alpha\delta V^{\rm proper}$ with $\delta V^{\rm proper}=\delta V_\infty/\alpha$), and inserting $\Delta\varphi(t)=\Delta\varphi_0+ (2e/\hbar)\delta V_\infty t$ into \eqref{4.12}, the time-averaged current near the edge of stability yields the operational critical value. The amplitude that multiplies the slowly varying $\sin\Delta\varphi$ is $\alpha I_c^{\rm proper}$, while the slope that ties phase evolution to the asymptotic control $\delta V_\infty$ is the flat-space value $2e/\hbar$ (because it is already written in terms of asymptotic quantities). As a result, the critical current inferred from an asymptotic biasing protocol scales as
\begin{equation}
I_{c,\infty}^{\rm (bias)} \simeq \alpha I_c^{\rm proper},
\label{4.16}
\end{equation}
for the fixed asymptotic control. Equations \eqref{4.13}-\eqref{4.16} make explicit that both operational notions of “critical current” (phase-clamped versus bias-inferred) agree on the one-power $\alpha$ scaling for static Schwarzschild backgrounds.

For completeness, it is useful to state an invariant version. Let $j^\mu$ be the conserved charge current and $\xi^\mu$ the timelike Killing field. The flux per unit Killing time through $S$ is $\displaystyle I_\infty=\int_S \alpha j^\mu d\Sigma_\mu$. The control that keeps the phase stationary is encoded in the electrochemical potential $\tilde{\mu}=\mu-q\Phi$, whose redshift $\alpha\tilde{\mu}$ is fixed on each bank in equilibrium and whose small DC imbalance generated at infinity determines the phase-slip rate by \eqref{2.18}. The product of these two projections (one on the flux side and one on the control side) underlies the single-$\alpha$ law for asymptotic currents in static backgrounds:
\begin{equation}
I_{c,\infty}^{\rm (bias)} = \alpha I_c^{\rm proper}[1+\mathcal{O}((L/R_c)^2)].
\label{4.17}
\end{equation}
In Subsection \ref{ssec4.3} we analyze the weak-field and near-horizon limits of these relations and show how the single-$\alpha$ scaling manifests in those regimes.

\subsection{Near-horizon and weak-field limits} \label{ssec4.3}
The relations \eqref{4.12}-\eqref{4.17} admit simple asymptotics in two opposite regimes. We record the leading scalings and the size of geometric corrections, emphasizing the distinction between phase-clamped and bias-inferred critical currents. Throughout, we keep the sign convention from Section \ref{sec3}: the Newtonian potential $\Phi_N$ is negative in an attractive field and the Schwarzschild lapse expands as $\alpha(r)\simeq 1+\Phi_N(r)/c^2$ with $|\Phi_N|\ll c^2$.

For the near-horizon (Rindler) scaling, let $r=r_s+\delta r$ with $r_s=2GM/c^2$ and $\delta r\ll r_s$. Along a static radial ray, the proper distance to the horizon is $\rho$, and the lapse behaves as
\begin{equation}
\alpha(r)\simeq \kappa \rho,
\qquad
\kappa=\frac{c^4}{4GM}
\label{4.18}
\end{equation}
(the surface gravity). For a junction localized within a region of size $L\ll \rho$, both $\alpha$ and $\gamma_{ij}$ are approximately constant across the device up to $\mathcal{O}((L/\rho)^2)$.

With the local current-phase law \eqref{4.8} fixed, the asymptotic flux per unit Killing time in a phase-clamped protocol follows from \eqref{4.12}:
\begin{equation}
I_{c,\infty}^{\rm (phase)}\simeq \alpha I_c^{\rm proper}
\simeq (\kappa\rho)I_c^{\rm proper}.
\label{4.19}
\end{equation}
Hence, the measurable DC critical current vanishes linearly with $\rho$ as the junction approaches the horizon, even when the local constitutive physics is unchanged. In a bias-inferred protocol the same single-$\alpha$ scaling applies (Sec. \ref{ssec4.2}):
\begin{equation}
I_{c,\infty}^{\rm (bias)}\simeq \alpha I_c^{\rm proper}
\simeq (\kappa\rho)I_c^{\rm proper}.
\label{4.20}
\end{equation}
Equations \eqref{4.19}-\eqref{4.20} are insensitive to microscopic details of the barrier, which enter only through $I_c^{\rm proper}$.

\begin{figure}
    \centering
    \includegraphics[width=\columnwidth]{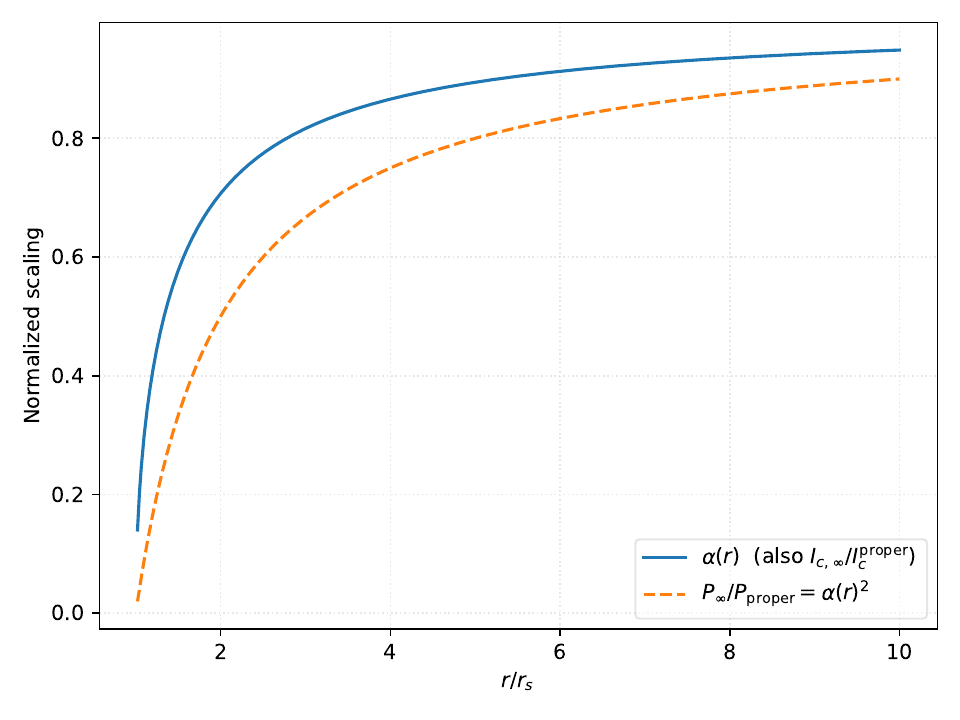}
    \caption{Lapse-radius profile and near-horizon scalings. In Schwarzschild, \(\alpha(r)=\sqrt{1-r_s/r}\) governs the redshift. The DC critical current observed at infinity scales with a single power, \(I_{c,\infty}/I_c^{\rm proper}=\alpha(r)\), while the power scales quadratically, \(P_\infty/P_{\rm proper}=\alpha(r)^2\).}
    \label{fig3}
\end{figure}

In Figure \ref{fig3}, the kinematic separation between flux-like and energy-like observables is shown. The current measured in Killing time is the proper current diluted by \(d\tau/dt=\alpha\), hence \(I_{c,\infty}=\alpha I_c^{\rm proper}\). Power acquires two factors of \(\alpha\): one from the rate and one from the redshift of energy per pair, giving \(P_\infty=\alpha^2 P_{\rm proper}\). As \(r\to r_s\), \(\alpha\to 0\) and the current seen at infinity vanishes linearly, while power, vanishes quadratically. In the Rindler limit with proper distance \(\rho\) from the horizon, \(\alpha\simeq \kappa\rho\) so \(I_{c,\infty}\propto \rho\) and \(P_\infty\propto \rho^2\). This linear-quadratic contrast is the corrected DC transport law and will anchor the near-horizon analysis and scaling arguments used later in this paper.

Two small parameters set the accuracy of these scalings. First, geometric variations across the device give relative corrections $\mathcal{O}((L/\rho)^2)$. Second, background curvature across the device produces corrections $\mathcal{O}((L/R_c)^2)$, where one may estimate $R_c$ from the Kretschmann scalar in Schwarzschild. Using $K=48G^2M^2/(c^4 r^6)$, a convenient curvature radius is $R_c\sim K^{-1/2}\sim r^3/(\sqrt{48}GM/c^2)$. Near the horizon $r\sim r_s$, so $R_c\sim r_s/\sqrt{48}$; our short-junction assumption requires $L\ll R_c$.

For the weak-field (Newtonian) limit, let $|\Phi_N|\ll c^2$ with $\alpha(r)\simeq 1+\Phi_N(r)/c^2$. For a compact junction at radius $r$, one may take $\alpha\simeq \alpha(r)$ across its area. Equations \eqref{4.13} and \eqref{4.16} then give, to first order,
\begin{align}
I_{c,\infty}^{\rm (phase)} &\simeq \left(1+\frac{\Phi_N}{c^2}\right) I_c^{\rm proper}, \nonumber \\
I_{c,\infty}^{\rm (bias)} &\simeq \left(1+\frac{\Phi_N}{c^2}\right) I_c^{\rm proper}.
\label{4.21}
\end{align}
Thus, the fractional gravitational correction to the DC critical current is the same in the two protocols in a static background: it carries a single power of $\alpha$ because the phase-locking condition \eqref{4.14} removes any extra redshift from the control side when written in asymptotic variables. If the device spans a small vertical extent $\Delta h$ in an approximately uniform field $g$, one may replace the potential difference by $\Delta\Phi_N\simeq g \Delta h$ when comparing two otherwise identical junctions placed at different heights. Tidal corrections remain $\mathcal{O}((L/R_c)^2)$ with $R_c$ set by the local Schwarzschild curvature scale.

In the left panel of Figure \ref{fig7,8}, two universal behaviors control the near-horizon regime. First, radial null propagation accumulates a Shapiro phase \(\psi=\Omega_\infty\int dt\) with \(dt/dr=(1-r_s/r)^{-1}\), yielding the familiar logarithm \(\psi_{\rm Shapiro}\sim \Omega_\infty r_s\ln[(R-r_s)/(r-r_s)]\). Plotting against \(\ln(r-r_s)\) exposes the straight-line trend and clarifies that the divergence is only logarithmic. Second, the DC transport read out at infinity is governed by the lapse: \(I_{c,\infty}=\alpha I_c^{\rm proper}\), so near the horizon \(\alpha \simeq \kappa \rho\) with \(\kappa=1/(2r_s)\) and \(\rho\) the proper distance, giving a linear vanishing \(I_{c,\infty}\propto \rho\). The single panel juxtaposes these scalings by labeling the abscissa with \(\ln(r-r_s)\) (bottom) and the corresponding \(\rho\) (top). Together, they summarize the kinematics relevant for the Rindler limit: logarithmically increasing propagation phase alongside linearly decreasing observable current. Both the AC frequency under proper bias and the DC critical current observed at infinity scale with the lapse \(\alpha(r)\) (see the right panel of Figure \ref{fig7,8}). The weak-field expansion at \(r\gg r_s\) gives \(\alpha_{\rm lin}=1-\frac{r_s}{2r}+ \mathcal{O}((r_s/r)^2)\). The figure demonstrates that the first-order approximation is already accurate to the \(2.06\%\) level at \(r=3r_s\) and improves monotonically with radius (e.g., \(<1\%\) by \(r\approx6r_s\), \(<0.2\%\) by \(r\approx10r_s)\). This validates using the linearized formulas in Section \ref{sec4} for analytic intuition and quick estimates, while the exact \(\alpha(r)\) should be retained for near-horizon analyses and for quantitative comparisons across a wide radial range.
\begin{figure}
    \centering
    \includegraphics[width=\columnwidth]{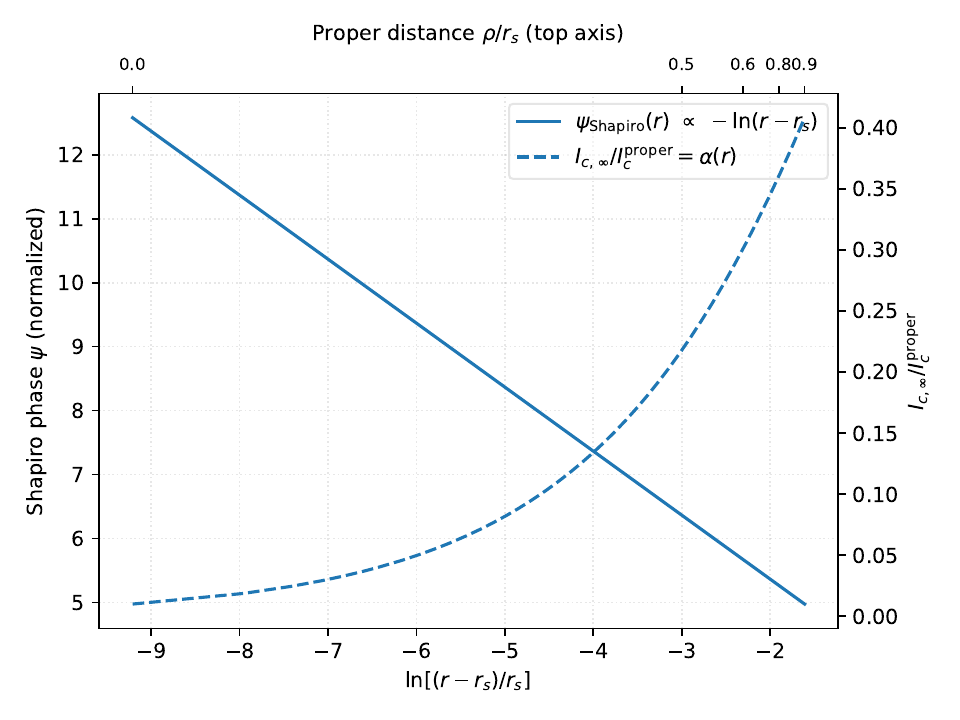}
    \includegraphics[width=\columnwidth]{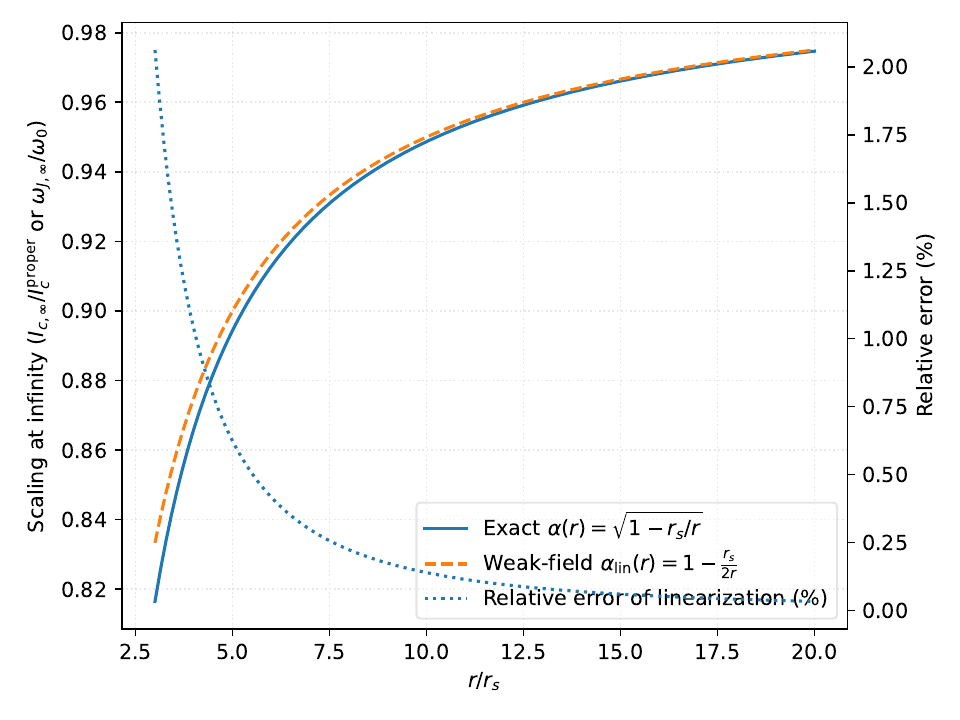}
    \caption{Left panel: Near-horizon behavior. The Shapiro phase grows logarithmically as the horizon is approached: \(\psi_{\rm Shapiro}(r)\propto -\ln(r-r_s)\) when plotted against \(\ln[(r-r_s)/r_s]\) (solid). On the same panel, the DC critical current measured at infinity vanishes with a single power of the lapse, \(I_{c,\infty}/I_c^{\rm proper}=\alpha(r)\) (dashed). The top axis shows the corresponding proper distance \(\rho/r_s\) from the horizon. Right panel: Weak-field linearization check. Exact single-\(\alpha\) scaling \(\alpha(r)=\sqrt{1-r_s/r}\) (solid) against its first-order expansion \(\alpha_{\rm lin}(r)=1-\frac{r_s}{2r}\) (dashed) over \(3\le r/r_s\le 20\). The dotted curve shows the relative error \((\alpha_{\rm lin}-\alpha)/\alpha\) in percent.}
    \label{fig7,8}
\end{figure}

We now check consistency with AC scaling. The near-horizon trend in \eqref{4.19}-\eqref{4.20} is complementary to the AC result \eqref{3.6}: for locally imposed biases, $\dot{\Delta\varphi}_\infty\propto \alpha$ is suppressed as $\rho\to 0$, while for bias specified at infinity the frequency is independent of $\alpha$ [Eq. \eqref{3.3}]. In the weak field, the same consistency appears between \eqref{3.4}-\eqref{3.5} and \eqref{4.21}: the frequency shift is $\mathcal{O}(\Phi_N/c^2)$, and the DC critical current picks up a commensurate $\mathcal{O}(\Phi_N/c^2)$ correction with the same coefficient in both phase-clamped and bias-inferred protocols. These parallel scalings will be useful in Section \ref{sec5} when analyzing interference in a vertical SQUID.

\section{The vertical SQUID in Schwarzschild} \label{sec5}
We consider a superconducting loop that contains two weak links located at radii $r_1$ and $r_2=r_1+\Delta r$ in the Schwarzschild exterior. The loop lies on a static slice $\Sigma_t$, and its plane is arranged so that the two junctions are vertically separated while the rest of the loop remains approximately isopotential with respect to the static observers. The aim is twofold. First, we formulate the gauge-invariant phase dynamics of the two junctions, including the proper-asymptotic dictionary from Section \ref{sec2}. Second, we impose the loop constraint from fluxoid quantization on $\Sigma_t$, with a self-inductance that can be finite. This subsection sets up the coupled equations that determine the interference pattern and prepares the ground for the weak-field analysis in Subsection \ref{ssec5.2}.

\subsection{Phase dynamics and loop constraint} \label{ssec5.1}
Let $\varphi_i(t)$ be the gauge-invariant phase drop across junction $i=1,2$, defined on $\Sigma_t$ by an integral of the form in \eqref{2.4} evaluated on a short path that threads the corresponding weak link. The junctions are static with respect to the Killing congruence, and the local current-phase relations on $\Sigma_t$ are
\begin{equation}
I_i(t) = I_{ci}^{\rm proper} \sin\varphi_i(t),
\label{5.1}
\end{equation}
with $I_{ci}^{\rm proper}$ computed from the local GL analysis of Section \ref{sec4}. The currents that an observer at infinity infers per unit Killing time obey
\begin{equation}
I_{i,\infty}(t)\simeq \alpha_i I_{ci}^{\rm proper} \sin\varphi_i(t),
\qquad \alpha_i\equiv \alpha(r_i),
\label{5.2}
\end{equation}
in the short-junction, weak-curvature regime. Equation \eqref{5.2} follows from the hypersurface flux mapping \eqref{2.19}-\eqref{2.20} and will be used only when converting final answers to asymptotic observables.

The phase evolution of each junction is fixed by the redshifted Josephson relation. In the presence of small, possibly time-dependent electrochemical drops applied in the local static frames,
\begin{equation}
\dot{\varphi}_i(t)=\frac{2e}{\hbar} \alpha_i V_i^{\rm proper}(t),
\label{5.3}
\end{equation}
with $V_i^{\rm proper}$ defined as in Subsection \ref{ssec2.1}. When an RF drive is generated at infinity with angular frequency $\Omega_\infty$, the proper drive seen at $r_i$ is $V_i^{\rm proper}(t)=\tfrac{1}{\alpha_i}\text{Re}\big\{\kappa_i V_{\rm rf}^\infty e^{,i(\Omega_\infty t-\psi_i)}\big\}$ plus any DC component, as in \eqref{3.7}. Substituting into \eqref{5.3} shows that the AC part of $\dot\varphi_i$ depends on the asymptotic amplitude $V_{\rm rf}^\infty$ but is independent of $\alpha_i$, while the DC part carries the factor $\alpha_i$ through the locally imposed bias, which is fully consistent with \eqref{3.2}-\eqref{3.3}.

The two phases are not independent. On the static slice $\Sigma_t$, fluxoid quantization around the loop gives
\begin{equation}
\varphi_1-\varphi_2
= 2\pi \frac{\Phi_{\rm tot}}{\Phi_0},
\qquad \Phi_0=\frac{h}{2e},
\label{5.4}
\end{equation}
where $\Phi_{\rm tot}$ is the magnetic flux threading any surface bounded by the loop on $\Sigma_t$. If the loop has self-inductance $L$ (defined with respect to proper fields on $\Sigma_t$), then
\begin{equation}
\Phi_{\rm tot}=\Phi_{\rm ext}+L I_{\rm circ},
\label{5.5}
\end{equation}
with $\Phi_{\rm ext}$ the externally applied flux and $I_{\rm circ}$ the circulating current on the slice. A convenient parametrization assigns branch currents
\begin{equation}
I_1=\frac{I_b}{2}+I_{\rm circ},\qquad
I_2=\frac{I_b}{2}-I_{\rm circ},
\label{5.6}
\end{equation}
where $I_b$ is the loop bias on $\Sigma_t$. In a stationary DC state, the junction currents equal the branch currents,
\begin{equation}
I_{ci}^{\rm proper} \sin\varphi_i = I_i,
\label{5.7}
\end{equation}
and the constraint \eqref{5.4}-\eqref{5.5} couples $\varphi_1$ and $\varphi_2$ through $I_{\rm circ}$.

Equations \eqref{5.4}-\eqref{5.7} form a closed system for $(\varphi_1,\varphi_2,I_{\rm circ})$ given $(I_b,\Phi_{\rm ext})$ on $\Sigma_t$. In the limit $L\to 0$, the quantization condition reduces to
\begin{equation}
\varphi_1-\varphi_2 = 2\pi \frac{\Phi_{\rm ext}}{\Phi_0},
\label{5.8}
\end{equation}
where $\Phi_0=h/q$ with $q=2e$. Here, the loop realizes the familiar algebraic interference. For identical junctions $(I_{c1}^{\rm proper}=I_{c2}^{\rm proper}=I_c^{\rm proper})$ and $L=0$ one may eliminate $\varphi_{1,2}$ to obtain the static critical current on $\Sigma_t$,
\begin{equation}
I_{c,\rm loop}^{\rm proper}(\Phi_{\rm ext})
= 2 I_c^{\rm proper}\big|\cos(\pi \Phi_{\rm ext}/\Phi_0)\big|.
\label{5.9}
\end{equation}
When $L$ is finite, the implicit equation \eqref{5.4}-\eqref{5.7} yields the standard dc-SQUID response with parameter $\beta_L\equiv 2\pi L I_c^{\rm proper}/\Phi_0$, now understood intrinsically on $\Sigma_t$. The conversion to asymptotic observables multiplies currents by $\alpha$ according to \eqref{5.2} and leaves $\Phi_{\rm ext}$ unchanged, since magnetic flux is defined purely on $\Sigma_t$.

Time-dependent operation follows by differentiating \eqref{5.4}. Using \eqref{5.3} and writing $\Phi_{\rm tot}=\Phi_{\rm ext}(t)+L I_{\rm circ}(t)$,
\begin{equation}
\dot{\varphi}_1-\dot{\varphi}_2
= 2\pi \frac{\dot{\Phi}_{\rm ext}+L \dot{I}_{\rm circ}}{\Phi_0}
= \frac{2e}{\hbar}\left(\alpha_1 V_1^{\rm proper}
- \alpha_2 V_2^{\rm proper}\right).
  \label{5.10}
\end{equation}
Equation \eqref{5.10} expresses a balance: differential phase driving from biases and propagation phases equals the electromotive contribution from a changing loop flux and from the inductive response. For a steady RF drive from infinity with frequency $\Omega_\infty$, one may insert the explicit forms of $V_i^{\rm proper}(t)$ and $\Phi_{\rm ext}(t)$ to obtain phase locking conditions that generalize \eqref{3.10} to the two-junction case. The lobe shifts and envelopes that result will be organized in Subsection \ref{ssec5.2}, where we specialize to weak fields and extract compact expressions for the vertical shift induced by $\alpha_1\neq \alpha_2$.

\subsection{Weak-field interference shift} \label{ssec5.2}
We work in the weak-field domain $|\Phi_N|\ll c^2$ with $\alpha(r)\simeq 1+\Phi_N(r)/c^2$ and $\Phi_N=-GM/(rc^2)$. The two junctions occupy radii $r_1$ and $r_2=r_1+\Delta r$, so that $\alpha_1=\alpha(r_1)$ and $\alpha_2=\alpha(r_2)$ differ by
\begin{equation}
\Delta\alpha \equiv \alpha_1-\alpha_2 \simeq \frac{\Phi_N(r_1)-\Phi_N(r_2)}{c^2}\simeq \frac{g \Delta h}{c^2},
\label{5.11}
\end{equation}
where $g$ is the local gravitational acceleration and $\Delta h$ the vertical separation measured on $\Sigma_t$.

We first analyze the purely DC, $L\to 0$ loop in the stationary regime, then turn to the RF-locked case. The purpose is to isolate whether gravity shifts the interference pattern in external flux, or merely distorts its envelope, and to identify which effects are truly due to redshift rather than kinematic propagation.

For a DC operation with no RF drive, let $I_{c1}^{\rm proper}=I_{c2}^{\rm proper}\equiv I_c^{\rm proper}$ and $L=0$. On $\Sigma_t$ the branch relations and the loop constraint give, as in \eqref{5.1} and \eqref{5.8},
\begin{equation}
I_1=I_c^{\rm proper}\sin\varphi_1,\qquad
I_2=I_c^{\rm proper}\sin\varphi_2,\qquad
\end{equation}
\begin{equation}
\varphi_1-\varphi_2=2\pi f,\quad f\equiv \frac{\Phi_{\rm ext}}{\Phi_0}.
\label{5.12}
\end{equation}
In the flat-space limit, the envelope reduces to classic Josephson/SQUID interferometry \cite{Jaklevic:1964ysq,tinkham1975introduction}. In gravitational settings, phase shifts in superfluid/Josephson interferometers have been analyzed in the Newtonian limit \cite{Anandan:1981zd,Anandan_1984}.
 Define $\chi=(\varphi_1+\varphi_2)/2$. The total current inferred at infinity is
\begin{equation}
I_\infty(f,\chi)=\alpha_1 I_c^{\rm proper}\sin(\chi+\pi f)
+\alpha_2 I_c^{\rm proper}\sin(\chi-\pi f).
\label{5.13}
\end{equation}
Maximizing over $\chi$ at fixed $f$ yields the interference envelope
\begin{equation}
I_{\infty,\max}(f)=2\bar{\alpha} I_c^{\rm proper}
\sqrt{ \cos^{2}(\pi f)+\rho^{2}\sin^{2}(\pi f) },\end{equation}
\begin{equation}
\bar{\alpha}\equiv \frac{\alpha_1+\alpha_2}{2},\quad
\rho\equiv \frac{\alpha_1-\alpha_2}{\alpha_1+\alpha_2}.
\label{5.14}
\end{equation}
Equation \eqref{5.14} is the weak-field “vertical SQUID” envelope to leading order in junction shortness and curvature.

Two conclusions follow immediately. First, there is no linear-in-$\rho$ shift of the lobe centers in $f$. The maxima remain near integer $f$ and the minima near half-integer $f$; gravity does not translate the pattern in $\Phi_{\rm ext}$ at $\mathcal{O}(\rho)$. Second, redshift produces a weak distortion of the cosine envelope: for $|\rho|\ll 1$,
\begin{equation}
I_{\infty,\max}(f)\simeq 2\bar{\alpha} I_c^{\rm proper}
\left[|\cos(\pi f)|
+\frac{\rho^{2}}{2}\frac{\sin^{2}(\pi f)}{|\cos(\pi f)|}\right],
\label{5.15}
\end{equation}
a quadratic-in-$\ rho$ effect. Using \eqref{5.11},
\begin{equation}
\rho \simeq \frac{\Delta\alpha}{2\bar{\alpha}}
\simeq \frac{g \Delta h}{2 c^{2}},
\label{5.16}
\end{equation}
so the fractional distortion is $\mathcal{O}\left((g \Delta h/c^{2})^{2}\right)$. Thus, in DC operation without RF, the weak gravitational field does not shift the interference in flux; it slightly deforms the lobe shape and rescales the overall amplitude by $2\bar{\alpha}$.

A finite inductance $L$ produces the familiar self-biasing of the loop. In the weak-inductance regime $\beta_L\equiv 2\pi L I_c^{\rm proper}/\Phi_0\ll 1$, the leading effect is to renormalize the envelope curvature near its extrema without generating an $\mathcal{O}(\rho)$ flux shift. Corrections scale as $\beta_L$ and $\rho^2\beta_L$ and are subleading under our assumptions.

In RF-locked operation, consider now a monochromatic drive generated at infinity with angular frequency $\Omega_\infty$ and no DC bias from infinity. Each junction phase obeys \eqref{5.3} with
\begin{equation}
V_i^{\rm proper}(t)=\tfrac{1}{\alpha_i}\text{Re}\big\{\kappa_i V_{\rm rf}^{\infty}e^{i(\Omega_\infty t-\psi_i)}\big\}.
\end{equation}
Integrating and applying the Jacobi-Anger expansion as in Section \ref{sec3} yields, on the $n$-th step, an averaged response of the form
\begin{equation}
\langle I_\infty\rangle
=\alpha_1 I_c^{\rm proper} J_n(a_1)\sin(\phi_{01}-n\psi_1)
\end{equation}\begin{equation}+\alpha_2 I_c^{\rm proper} J_n(a_2)\sin(\phi_{02}-n\psi_2),
\label{5.17}
\end{equation}
where $a_i=(2e/\hbar) |V_{\rm rf,eff}^{\infty}|/\Omega_\infty$ projected onto junction $i$ and $\psi_i$ are the propagation phases from infinity to $r_i$. Eliminating $\phi_{0i}$ with the constraint $\phi_{01}-\phi_{02}=2\pi f$ and optimizing over the common offset gives an RF interference envelope analogous to \eqref{5.14},
\begin{equation}
\langle I_\infty\rangle_{\max}
= 2\bar{\alpha} I_c^{\rm proper}
\sqrt{\zeta },
\label{5.18}
\end{equation}
\begin{align}
\zeta&=\left[J_n(\bar{a})\cos\left(\pi f-\frac{n}{2}\Delta\psi\right)\right]^2 \nonumber \\
&+\rho^2 \left[J_n(\bar{a})\sin\left(\pi f-\frac{n}{2}\Delta\psi\right)\right]^2
\end{align}
with $\bar{a}=\tfrac{1}{2}(a_1+a_2)$, $\delta a=a_1-a_2$, and $\Delta\psi\equiv \psi_1-\psi_2$. With symmetric RF coupling from infinity ($a_1=a_2$), redshift alone does not produce a differential modulation; any apparent lobe translation arises solely from propagation phases $\Delta\psi$ (geometric $+$ Shapiro).

Two points are crucial: First, for symmetric coupling $\kappa_1=\kappa_2$ and short junctions, one has $a_1=a_2$ at leading order even though $\alpha_1\neq\alpha_2$, because the $\alpha_i$ that multiplies the proper drive cancels the $\alpha_i^{-1}$ in \eqref{5.3}. Hence, the redshift does not by itself create a differential RF modulation. Second, \eqref{5.18} shows an apparent flux shift
\begin{equation}
\Delta \Phi_{\rm RF}
= \frac{\Phi_0}{2\pi} n\Delta\psi ,
\label{5.19}
\end{equation}
i.e. a translation $f\mapsto f-\tfrac{n}{2\pi}\Delta\psi$ of the lobe pattern on the $n$-th step \cite{Shapiro:1963nhj}. This shift is entirely controlled by the differential propagation phase \cite{Shapiro:1964uw,Possel:2019zbu}. In a static Schwarzschild background, there is no Sagnac term; $\Delta\psi$ consists of the geometric path phase plus the (small) Shapiro delay. At weak field, $\Delta\psi= \Omega_\infty\left(\Delta t_{\rm geo}+\Delta t_{\rm Shapiro}\right)=\mathcal{O}(|\Phi_N|/c^2)$. Crucially, gravitational redshift per se does not contribute to $\Delta\psi$ under symmetric coupling, so the leading RF shift \eqref{5.19} is kinematic rather than redshift-driven. Any apparent lobe translation is set by propagation phases (geometric plus the Shapiro delay \cite{Shapiro:1964uw,Possel:2019zbu}) rather than by redshift itself when both bias and drive are specified at infinity.

If the couplings are intentionally imbalanced ($\kappa_1\neq \kappa_2$), or if the drive is applied locally in the proper frames rather than from infinity, then $a_1\neq a_2$ at $\mathcal{O}(\rho)$ and an additional, genuinely redshift-induced phase offset appears through the standard $J_n$-weighted mixing. To first order in both $\rho$ and $\delta a/\bar{a}$, the resulting shift of the interference phase is
\begin{equation}
\delta(\pi f)\big|_{\rm redshift}
\simeq \frac{1}{2}\frac{\delta a}{\bar{a}}\rho\tan\left(\pi f-\frac{n}{2}\Delta\psi\right),
\label{5.20}
\end{equation}
which is parametrically small and depends on the operating point within a lobe.

\begin{figure}
    \centering
    \includegraphics[width=\columnwidth]{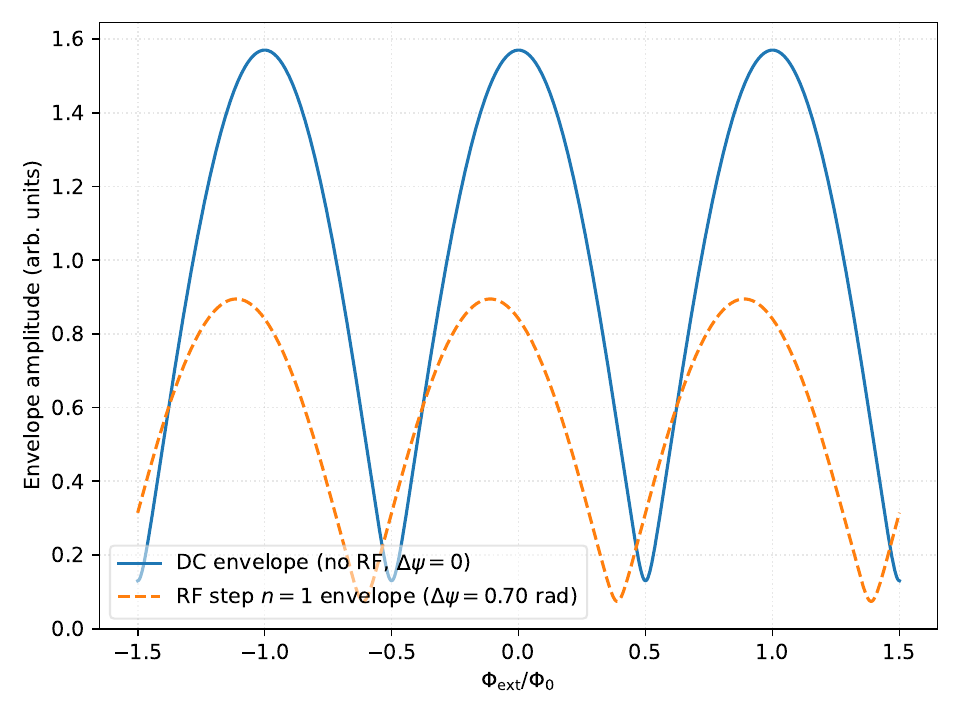}
    \caption{Vertical SQUID interference envelopes. DC pattern (solid) computed from two junctions at different radii with lapses \(\alpha_1=0.72\) and \(\alpha_2=0.85\) shows lobe centers at \(\Phi_{\rm ext}/\Phi_0 \in \mathbb{Z}\) with lifted minima due to the asymmetry but no shift in the centers. Under RF drive, the \(n=1\) Shapiro-step envelope (dashed) is translated by a propagation-phase difference \(\Delta\psi=0.70\ \mathrm{rad}\); the translation is set by \(\Delta\psi\), not by \(\Delta\alpha\).}
    \label{fig4}
\end{figure}

We model the vertical SQUID by summing the arm currents in the Killing frame (see Figure \ref{fig4}). For DC, the critical current observed at infinity follows
\begin{equation}
I_{\rm DC}(\Phi)=I_0\sqrt{\alpha_1^2+\alpha_2^2+2\alpha_1\alpha_2\cos\delta},\quad \delta=2\pi\Phi_{\rm ext}/\Phi_0,
\end{equation}
so, mild redshift imbalance only lifts the minima through the \(\alpha_1\alpha_2\) cross term while keeping the lobe centers at integer flux. This confirms that there is no \(\mathcal{O}(\Delta\alpha)\) translation of the DC interference pattern. With an RF drive defined at infinity, the amplitude of the \(n\)-th step inherits an extra propagation phase, giving a replacement \(\delta\to\delta+\Delta\psi\) in the envelope; the overall scale is set by \(|J_n(a)|\). The dashed curve (for \(n=1\)) shows that the apparent lobe translation arises from \(\Delta\psi\) alone. Thus, the figure encapsulates the separation of roles: redshift rescales amplitudes, while gravitational propagation modifies phases and shifts RF interference envelopes.

\subsection{Parameter ranges and observability} \label{ssec5.3}
The formulas above translate to asymptotic observables through two rules consistent with Section \ref{sec2}: (i) currents measured per unit Killing time acquire one power of the lapse relative to their proper values, \(I_\infty\simeq \alpha I_{\rm proper}\); (ii) the phase-bias conversion obeys \(\dot\theta=(q/\hbar) V^\infty\) when the control is specified at infinity, and \(\dot\theta=(q/\hbar) \alpha V^{\rm proper}\) when the control is local. Thus there is no additional lapse factor hidden in the phase-bias relation when everything is written in asymptotic variables. We now ask whether the redshift-dependent signatures are within reach in weak fields, or only in compact-object exteriors. Short-junction and weak-curvature assumptions are retained throughout.

We begin with the geometric constraints that validate the “local” treatment on a static slice. Let \(L\) be the proper thickness across the barrier and \(A\) the proper cross-section (both on \(\Sigma_t)\). The short-junction approximation requires \(L\ll\xi_{L R}\) in the banks and \(L\ll\xi_N\) in the barrier. Curvature and lapse inhomogeneity must be small across the device: \(L\ll R_c\) and \(L\ll \ell_\alpha\), where \(R_c\) is a curvature radius extracted from \(\gamma_{ij}\) and \(\ell_\alpha\equiv \alpha/|\nabla\alpha|\) sets the redshift variation scale along the transport direction. Under these conditions, the mapping from the proper GL solution to asymptotic fluxes (Sec. \ref{sec4}) holds to relative accuracy \( \mathcal{O}((L/R_c)^2+(L/\ell_\alpha)^2)\).

In a terrestrial weak field we adopt the consistent convention \(\alpha(r)\simeq 1+\Phi_N(r)/c^2\) with \(\Phi_N<0\) in an attractive potential and \(|\Phi_N|\ll c^2\). For a vertical separation \(\Delta h\) between the two junctions, the redshift contrast is \(\Delta\alpha\simeq [\Phi_N(r_1)-\Phi_N(r_2)]/c^2\simeq g \Delta h/c^2\). In DC operation without RF, the loop envelope inferred at infinity is given by \eqref{5.14} with \(\rho\simeq \Delta\alpha/(2\bar\alpha)\simeq g \Delta h/(2c^2)\). The lobe centers remain at integer and half-integer external flux; gravity does not translate the pattern in \(\Phi_{\rm ext}\) at \( \mathcal{O}(\rho)\). The envelope distortion is quadratic in \(\rho\), hence \( \mathcal{O}((g \Delta h/c^2)^2)\). For \(\Delta h\sim\) meters, this is far below practical detection. The overall amplitude rescales by \(2\bar\alpha\), but \(\bar\alpha=1+ \mathcal{O}(\Phi_N/c^2)\) differs from unity only at the \(10^{-9}-10^{-10}\) level over laboratory heights.

Under RF operation, the step loci in asymptotic variables remain fixed at \(\langle V_\infty\rangle=n\hbar\Omega_\infty/(2e)\) when both bias and drive are delivered from infinity (Sec. \ref{sec3}). The key observable is the lobe translation \eqref{5.19}, \(\Delta\Phi_{\rm RF}=(\Phi_0/2\pi),n,\Delta\psi\). The differential phase is \(\Delta\psi\simeq \Omega_\infty(\Delta t_{\rm geo}+\Delta t_{\rm Shapiro})\): the geometric part scales as \(\Omega_\infty,\Delta h/c\) and can be sizeable at microwave frequencies even for meter-scale \(\Delta h\); the Shapiro contribution is \( \mathcal{O}(\Phi_N/c^2)\). The geometric term is not a gravitational-redshift effect; it is a path-length difference and can be nulled by equalizing RF path lengths to the two junctions. With that cancellation, the residual \(\Delta\psi\) is the genuine gravitational (Shapiro) phase, hence \( \mathcal{O}(\Phi_N/c^2)\) and very small. Isolating it demands RF path matching and phase stability at parts in \(10^9\) over the integration time.

Compact-object exteriors present the opposite extreme. In Schwarzschild with \(\alpha\ll 1\) over macroscopic height scales, one can arrange \(\alpha_1\neq\alpha_2\) at \( \mathcal{O}(1)\) while still keeping \(L\ll R_c\) and \(L\ll \ell_\alpha\). Then the DC loop amplitude at infinity is suppressed by \(2\bar\alpha\), and both operational definitions of critical current scale with a single lapse: \(I_{c,\infty}^{\rm(phase)}\simeq \alpha I_c^{\rm proper}\) and \(I_{c,\infty}^{\rm(bias)}\simeq \alpha I_c^{\rm proper}\) [consistent with \eqref{4.19}-\eqref{4.20}]. The RF lobe translation remains governed by \(\Delta\psi\); near compact objects, the Shapiro component can compete with geometric phases over modest path differences because the logarithmic factor in the delay is enhanced by the strong potential. Redshift alone still does not translate the pattern in external flux; it reorganizes amplitudes and local proper frequencies (hence modulation indices) without moving the step loci expressed in asymptotic variables.

Finally, on inductance and asymmetries: a finite self-inductance \(L\) modifies the envelope but, in the weak-inductance regime \(\beta_L\ll 1\), does not generate an \( \mathcal{O}(\Delta\alpha)\) flux shift. Asymmetric junction parameters enter only through \(I_{c1,2}^{\rm proper}\) in the standard way, and conversion to asymptotic currents multiplies each by its local \(\alpha_i\). Asymmetric RF coupling \(\kappa_1\neq\kappa_2\) can generate an \( \mathcal{O}(\Delta\alpha)\) contribution via unequal modulation indices, leading to the small redshift-driven offset summarized in \eqref{5.20}. These effects are parametrically distinct from geometric propagation and from self-inductive feedback.

In brief, within the Schwarzschild-only analytic framework, gravitational redshift reorganizes amplitudes and proper frequencies but does not, by itself, translate the SQUID interference \cite{Jaklevic:1964ysq} pattern in external flux. In weak fields the redshift signatures are extremely small in DC operation and must be disentangled from propagation phases in RF operation; in strong fields the single-\( \alpha \) current scaling becomes the primary asymptotic signature, while RF lobe shifts can receive order-unity contributions from propagation phases in addition to any deliberately engineered redshift asymmetries. Geometric and analogue extensions include “geometric Josephson” junctions and circuit-gravity simulators \cite{Santos:2024cwf,Maceda:2025jti}, and AdS/BCFT generalizations in modified gravity \cite{Santos:2025ugv}.

\section{Conclusion} \label{sec6}
We have formulated Josephson physics in a static curved background in a way that makes the observable content manifest to an asymptotic observer. The central device has been the use of the timelike Killing field to define a redshift factor $\alpha$ and a foliation by static slices $\Sigma_t$. Within this kinematic structure, two gauge-invariant ingredients govern all results: the combination $\partial_\mu\theta-qA_\mu/\hbar$ that controls phase evolution, and the conserved current $j^\mu$ whose hypersurface flux encodes transport. The analysis has remained entirely local on $\Sigma_t$, so the passage from proper to asymptotic quantities is transparent and independent of coordinates.

The AC law derived in Section \ref{sec3} expresses the asymptotic phase rate as
\begin{equation}
\dot{\Delta\varphi}_\infty=\frac{2e}{\hbar} \left(\alpha_1 V_1^{\rm proper}-\alpha_2 V_2^{\rm proper}\right)
=\frac{2e}{\hbar} \left(V_1^\infty-V_2^\infty\right),
\end{equation}
i.e. with the $\alpha$ factors multiplying the proper drops, or, equivalently, using redshifted (asymptotic) drops $V_i^\infty\equiv\alpha_i V_i^{\rm proper}$. It reduces to the flat-space relation when the two banks share the same radius and makes explicit how a differential gravitational potential changes an otherwise identical local bias. The RF-driven problem shows that redshift alone does not displace the Shapiro-step loci if both the bias and the drive are specified at infinity; the positions remain fixed in terms of asymptotic voltages and the drive frequency. Propagation phases appear naturally and, in weak fields, account for the leading shift of the interference pattern under RF operation. Near a horizon, the proper frequency of an incoming drive is amplified, $\Omega_{\rm proper}=\Omega_\infty/\alpha$, yet the step loci in asymptotic variables remain pinned for the same operational reason.

The DC analysis of Section \ref{sec4} separates local constitutive physics from global readout. A short-junction solution on $\Sigma_t$ yields a proper current-phase relation; the flux measured per unit Killing time introduces one factor of $\alpha$ when converting to asymptotic current, $I_\infty\simeq \alpha I_{\rm proper}$. When the critical current is inferred from an asymptotic bias protocol that holds the phase stationary, the bias-to-phase conversion uses the asymptotic relation $\dot\theta=(2e/\hbar) V^\infty$, so the inferred critical current still scales with a single power of $\alpha$. By contrast, power picks up two powers: $P_\infty=I_\infty V_\infty\simeq \alpha^2 P_{\rm proper}$. These are kinematic statements that do not depend on microscopic details of the barrier and persist in both the weak-field and near-horizon limits, with controlled geometric corrections set by junction size compared to curvature and lapse-variation scales.

The vertical SQUID of Section \ref{sec5} illustrates how these principles interact in an interferometric setting. On a static slice, fluxoid quantization has its familiar form and is insensitive to $\alpha$. Consequently, in DC operation with identical junctions and negligible inductance, the interference pattern in external flux is not translated by gravity at linear order; instead, there is a small distortion of the envelope and an overall amplitude rescaling governed by $\bar\alpha$. Under RF drive from infinity, a translation appears if and only if a differential propagation phase is present. This distinction clarifies which features are genuinely gravitational (redshift in amplitudes and proper frequencies) and which are geometric (path and Shapiro phases). It also shows how to design protocols that either suppress propagation effects in order to test redshift alone, or exploit them as a sensitive interferometric resource.

The formalism relies on a hierarchy of scales that keeps the junction “short’’ relative to curvature and to variations of the lapse. Within that regime, replacing the standard flat-space GL functional by its intrinsic version on $\Sigma_t$ is consistent. The electromagnetic sector has been treated as a classical background, and backreaction on the metric has been neglected. These choices match the scope of the problem, which concerns the kinematics of phase and current in a fixed static geometry. 

Several extensions present themselves. The most immediate is to replace staticity by stationarity and analyze slowly rotating backgrounds. In Kerr, the shift vector is nonzero and introduces a gravitomagnetic contribution that couples directly to phase evolution. One expects new interference terms associated with frame dragging and a genuine Sagnac contribution to loop dynamics. The machinery needed for this generalization is close at hand: the (3+1) decomposition with lapse and shift, the projection of $j^\mu$ on non-orthogonal slices, and the corresponding generalization of the flux mapping.

A second extension concerns microscopic input. We have used a GL-level description on $\Sigma_t$, which suffices for tunneling and short weak links. It would be valuable to revisit the derivation starting from a curved-spacetime Bogoliubov-de Gennes or effective-action framework, with curvature couplings tracked explicitly in the coefficients. That exercise should leave the kinematic single-$\alpha$ and $\alpha^2$ (for power) factors intact, but it could refine how coherence lengths and interface parameters inherit curvature corrections.

A third direction involves composite circuits and multijunction networks in which different elements sit at different radii. The formalism immediately yields a rulebook for assembling such systems: local constitutive relations on $\Sigma_t$, loop constraints for each mesh, and a final projection to asymptotic observables. Networks that mix vertical and horizontal loops would allow one to isolate redshift from propagation by design, and to map out the transition from weak-field to near-horizon behavior within a single analytic scheme.

Finally, there is an avenue through analogue platforms. While our results target Schwarzschild as a background, the structure depends only on staticity. Engineered systems in which effective metrics or redshift profiles can be emulated provide testbeds for the kinematic aspects of the theory. Even without such emulation, conventional SQUID interferometry operated with carefully balanced RF delivery could separate propagation-induced phase shifts from true frequency-voltage relations, thereby confronting the predictions that are insensitive to material details.

To this end, the picture that emerges is coherent. Josephson phenomena in static curved spacetimes are determined by two invariant translations: proper-to-asymptotic phase rates and proper-to-asymptotic fluxes. Once those are in place, AC metrology, DC transport, and interferometric responses follow with minimal additional structure. The results are robust across limits and admit clear generalizations to stationary geometries. We expect that the same strategy can organize other phase-coherent transport effects, including superfluid weak links and hybrid circuits, and that it will remain compatible with more microscopic treatments that incorporate curvature at the level of the pairing theory. Our invariant dictionary, which are the proper \(\leftrightarrow\) asymptotic phases and fluxes, connects naturally to ongoing efforts that probe gravity with coherent quantum systems \cite{Gallerati:2022nwm,Bose:2024nhv}. On the metrology side, weak-field signatures align with redshift and AB-type tests in atom-based platforms \cite{Pumpo:2021sok,Overstreet:2021hea,Zheng:2022hwj,Chiao:2023ezj}. For superconducting circuits, gravity acts chiefly as a kinematic clock/energy reshaper; potential gravitational dephasing channels proposed for qubits \cite{Balatsky:2024tbv} and holographic realizations \cite{Horowitz:2011dz} offer complementary routes to explore beyond-flat-space Josephson physics.

\acknowledgements
R. P. and A. \"O. would like to acknowledge networking support of the COST Action CA21106 - COSMIC WISPers in the Dark Universe: Theory, astrophysics and experiments (CosmicWISPers), the COST Action CA22113 - Fundamental challenges in theoretical physics (THEORY-CHALLENGES), the COST Action CA21136 - Addressing observational tensions in cosmology with systematics and fundamental physics (CosmoVerse), the COST Action CA23130 - Bridging high and low energies in search of quantum gravity (BridgeQG), and the COST Action CA23115 - Relativistic Quantum Information (RQI) funded by COST (European Cooperation in Science and Technology). A. \"O. also thanks to EMU, TUBITAK, ULAKBIM (Turkiye) and SCOAP3 (Switzerland) for their support.

\appendix
\section{Gauge- and diffeomorphism-invariant definition of the phase drop across a weak link in a static geometry} \label{apdxA}
We work on a static spacetime $(\mathcal{M},g_{\mu\nu})$ admitting a timelike Killing field $\xi^\mu$ that is hypersurface-orthogonal. The orthogonal foliation by static slices $\Sigma_t$ has unit normal $n^\mu=\xi^\mu/\alpha$ and induced metric $\gamma_{ij}$. The superconducting condensate is a charged complex scalar $\psi=|\psi|e^{i\theta}$ of charge $q$, minimally coupled to the electromagnetic potential $A_\mu$. The basic gauge-invariant 1-form is (see \eqref{2.1})
\begin{equation}
p_\mu \equiv \hbar\,\partial_\mu\theta - qA_\mu,
\label{A.1}
\end{equation}
Throughout, the weak link is threaded by curves that lie inside a single static slice $\Sigma_t$. This is both natural (transport is read on $\Sigma_t$) and sufficient in a static spacetime, since no shift vector mixes the slice with time.

Let $x_1,x_2\in\Sigma_t$ be points in the two banks, chosen in a simply connected neighborhood of the junction and away from vortex cores. For any smooth curve $\mathcal{C}\subset\Sigma_t$ that threads the weak link once from $x_1$ to $x_2$, we define the gauge-invariant phase drop by (see \eqref{2.4})
\begin{equation}
\Delta\varphi[\mathcal{C}]
\equiv \frac{1}{\hbar}\int_{\mathcal{C}} \iota^{*}(p)
= \theta(x_2)-\theta(x_1)
-\frac{q}{\hbar}\int_{\mathcal{C}} A_i\,dx^i ,
\label{A.2}
\end{equation}
where $\iota^{*}$ denotes the pullback to $\Sigma_t$ and indices are raised/lowered with $\gamma_{ij}$ \cite{London1935,tinkham1975introduction} (See also \cite{Rastkhadiv:2025ibd}).

The definition \eqref{A.2} possesses the desired invariances and the correct topological dependence:
\begin{enumerate}
    \item Under $A_\mu\to A_\mu+\partial_\mu\chi$ and $\theta\to\theta+\frac{q}{\hbar}\chi$, the 1-form $p$ in \eqref{A.1} is invariant. Hence $\Delta\varphi[\mathcal{C}]$ is strictly gauge invariant.
    \item Let $\Phi:\Sigma_t\to\Sigma_t$ be a smooth spatial diffeomorphism (change of coordinates or a smooth deformation of the embedding). Since $\Delta\varphi[\mathcal{C}]=\hbar^{-1}\int_{\mathcal{C}}\iota^*(p)$, functoriality of the pullback gives $\Delta\varphi[\Phi(\mathcal{C})]=\hbar^{-1}\int_{\mathcal{C}}\Phi^{*}\iota^{*}(p)=\Delta\varphi[\mathcal{C}]$. Thus the construction depends only on the geometric image of the curve, not on coordinates.
    \item Consider two curves $\mathcal{C}_1,\mathcal{C}_2\subset\Sigma_t$ with the same endpoints $x_{1,2}$ and define the closed loop $\mathcal{L}=\mathcal{C}_1-\mathcal{C}_2$. In a vortex-free, simply connected domain one has $d(d\theta)=0$, so by Stokes’ theorem
    \begin{align} \label{A.3}
        &\Delta\varphi[\mathcal{C}_1]-\Delta\varphi[\mathcal{C}_2]
        = -\frac{q}{\hbar}\oint_{\mathcal{L}} A_i\,dx^i \nonumber \\
        &= -\frac{q}{\hbar}\int_{S(\mathcal{L})} B^i s_i\,dA 
        = -2\pi\,\frac{\Phi_{\rm loop}}{\Phi_0},
    \end{align}
    where $S(\mathcal{L})$ is any surface in $\Sigma_t$ with boundary $\mathcal{L}$, $\Phi_{\rm loop}=\int_{S}B^i s_i dA$ is the magnetic flux through $S$, and $\Phi_0=h/q$ is the flux quantum. If the region enclosed by $\mathcal{L}$ contains $N$ quantized vortices of the condensate, then $\oint_{\mathcal{L}} \partial_i\theta\,dx^i=2\pi N$, and \eqref{A.3} generalizes to the fluxoid form
    \begin{equation}
    \oint_{\mathcal{L}}\,\left(\partial_i\theta - \frac{q}{\hbar}A_i\right)dx^i
    = 2\pi N - 2\pi\,\frac{\Phi_{\rm loop}}{\Phi_0}.
    \label{A.4}
    \end{equation}
    Therefore, $\Delta\varphi[\mathcal{C}]$ is well defined modulo $2\pi$, and independent of the curve in a simply connected, vortex-free neighborhood. In a loop geometry, different choices of $\mathcal{C}$ are related precisely by \eqref{A.3}-\eqref{A.4}, which is the origin of the fluxoid quantization used in Section \ref{sec5}.
\end{enumerate}

It is useful to rephrase \eqref{A.2} in bundle-theoretic terms. Let $P\to\Sigma_t$ be the principal $U(1)$ bundle with connection $A$. The order parameter $\psi$ is a section of an associated complex line bundle. Parallel transport of $\psi(x_1)$ along $\mathcal{C}$ with respect to $A$ yields a complex number at $x_2$ proportional to $\exp\left(\frac{i q}{\hbar}\int_{\mathcal{C}}A\right)\psi(x_1)$. The gauge-invariant phase difference is the argument of the normalized inner product between the transported $\psi(x_1)$ and $\psi(x_2)$:
\begin{equation}
e^{\,i\Delta\varphi[\mathcal{C}]}
=\frac{\psi(x_2)}{|\psi(x_2)|}\,
\frac{\psi(x_1)^*}{|\psi(x_1)|}\,
\exp\,\left(-\frac{i q}{\hbar}\int_{\mathcal{C}}A\right).
\label{A.5}
\end{equation}
Taking the argument reproduces \eqref{A.2}. Equation \eqref{A.5} shows explicitly that $\Delta\varphi$ compares condensate phases at separated points only after the appropriate $U(1)$ parallel transport; this is the gauge-covariant content of the “phase difference.”

The definition is consistent with the kinematics on $\Sigma_t$ used in the main text. Projecting $p_\mu$ tangentially to $\Sigma_t$ gives (see \eqref{2.6})
\begin{equation}
p_i=\hbar\,\partial_i\theta - q A_i,
\label{A.6}
\end{equation}
so that $\hbar\,\Delta\varphi[\mathcal{C}]=\int_{\mathcal{C}} p_i\,dx^i$. Projecting $p_\mu$ on the static observers $U^\mu=n^\mu$ gives the local Josephson relation $U^\mu p_\mu=\mu$, cf. \eqref{2.5}, which underlies the phase-evolution statement \eqref{2.3}. These two projections are complementary: the spatial one encodes the instantaneous phase drop across the weak link; the temporal one ties phase evolution to electrochemical control.

Time dependence and Faraday induction enter in a manner that preserves gauge and diffeomorphism invariance. Differentiating \eqref{A.2} with respect to the Killing time $(t)$, and using the Maxwell-Faraday law on $\Sigma_t$ for the tangential electric field, one obtains the balance equation quoted in \eqref{2.9}:
\begin{equation}
\frac{d}{dt}\,\Delta\varphi
= \frac{q}{\hbar}\left(\alpha_1 V^{\rm proper}_1
-\alpha_2 V^{\rm proper}_2\right)
+ \frac{q}{\hbar}\,\frac{d}{dt}\left(\frac{\Phi_{\rm loop}}{\Phi_0}\right),
\label{A.7}
\end{equation}
with the second term absent for an isolated junction. The first term follows from the temporal projection of $p_\mu$ together with the Tolman-Ehrenfest redshift summarized in Subsection \ref{ssec2.1}; equivalently, the lapse $\alpha_i$ multiplies the proper drops $V^{\rm proper}_i$. The second term is the geometrical electromotive contribution from a changing flux on $\Sigma_t$. Both terms are intrinsically defined and remain invariant under gauge transformations and slice-preserving diffeomorphisms.

Finally, we note the practical choice of endpoints and curves. In a short junction, it is natural to pick $x_{1,2}$ on the two interfaces and to take $\mathcal{C}$ to cross the barrier along the unit normal to the interfaces. For $L\ll R_c$ and $L\ll \ell_\alpha$, the value of $\Delta\varphi$ is insensitive to the detailed placement of $x_{1,2}$ within the coherence-length boundary layers and to small deformations of $\mathcal{C}$, up to corrections $\mathcal{O}((L/R_c)^2)$. This justifies replacing the microscopic geometry by any convenient, smooth transversal when evaluating the current-phase relation on $\Sigma_t$, as done in Section \ref{sec4}.

Equations \eqref{A.2}-\eqref{A.7} provide a self-contained, gauge- and diffeomorphism-invariant notion of the phase drop across a weak link in a static geometry. They reduce to the familiar flat-space expressions in the limit $\alpha\to 1$, and they are the unique structures compatible with the covariant kinematics and conservation laws used throughout the main text.

\section{Tolman-Ehrenfest relations and electrochemical potential in superconductors} \label{apdxB}
We summarize the equilibrium redshift relations relevant for superconductors on a static spacetime $(\mathcal{M},g_{\mu\nu})$ that admits a hypersurface-orthogonal timelike Killing field $\xi^\mu$. Static observers have four-velocity $U^\mu=\xi^\mu/\alpha$, with lapse $\alpha=\sqrt{-\xi^\mu\xi_\mu}$. Their proper time is $d\tau=\alpha\,dt$, and spatial tensors live on the slices $\Sigma_t$ orthogonal to $U^\mu$.

The electromagnetic field is described by a potential $A_\mu$ with field strength $F_{\mu\nu}$. Static observers measure the scalar potential
\begin{equation}
\Phi \equiv -\,U^\mu A_\mu ,
\label{B.1}
\end{equation}
and the electric field $E_\mu=F_{\mu\nu}U^\nu$ tangent to $\Sigma_t$. The combinations $\Phi$ and $E_\mu$ are gauge invariant.

A normal relativistic fluid in static equilibrium satisfies the standard Tolman-Ehrenfest (TE) relations
\begin{equation}
T\,\alpha=\text{const},\qquad \mu\,\alpha=\text{const},
\label{B.2}
\end{equation}
where $T$ is the temperature and $\mu$ the chemical potential per particle measured in the local rest frame. For charged media, the operative variable entering transport is the electrochemical potential \cite{Tolman:1930ona,Tolman:1930zza,Lima:2019brf}
\begin{equation}
\tilde\mu \equiv \mu - q\,\Phi ,
\label{B.3}
\end{equation}
with $q$ the carrier charge. The static-equilibrium generalization of \eqref{B.2} is
\begin{equation}
\alpha\,\tilde\mu=\text{const on each connected equilibrium domain}.
\label{B.4}
\end{equation}
Equation \eqref{B.4} is the relation quoted as \eqref{2.7} in the main text. We sketch complementary derivations and note their role in superconducting kinematics.

Let us do a thermodynamic derivation. Consider a stationary configuration of a charged fluid on $\Sigma_t$. Let $u$ be the internal energy density and $n$ the particle density as measured by static observers. The local first law reads
\begin{equation}
du = T\,ds + \mu\,dn .
\label{B.5}
\end{equation}
Hydrostatic balance for a charged fluid is
\begin{equation}
\nabla_i p = -(u+p)\,\nabla_i\ln\alpha + n\,q\,E_i ,
\label{B.6}
\end{equation}
where $\nabla_i$ is the Levi-Civita connection of $\gamma_{ij}$. Substitute \eqref{B.6} into \eqref{B.7} $dp=s\,dT+n\,d\mu$ to obtain
\begin{equation}
s\,\nabla_i T + n\,\nabla_i \mu
= -(u+p)\,\nabla_i\ln\alpha + n\,q\,E_i .
\label{B.7}
\end{equation}
Using $u+p=Ts+\mu n$ from Euler’s relation and rearranging,
\begin{equation}
s\,\nabla_i(\alpha T) + n\,\nabla_i(\alpha\mu - \alpha q\,\Phi)=0 .
\label{B.8}
\end{equation}
In a connected equilibrium configuration without stationary heat or particle fluxes, both $s$ and $n$ are finite; hence, each bracket must be spatially constant. This yields
\begin{equation}
\alpha T=\text{const},\qquad \alpha(\mu-q\Phi)=\text{const},
\label{B.9}
\end{equation}
which reproduce \eqref{B.2} and \eqref{B.4}.

Let $\mathcal{E}$ be the Noether charge per particle associated with time translations generated by $\xi^\mu$. For a charged medium in static equilibrium one finds the constant of motion
\begin{equation}
\mathcal{E} \equiv -\,\xi^\mu p_\mu - q\,\xi^\mu A_\mu
= \alpha\,(\mu - q\,\Phi)
= \alpha\,\tilde\mu ,
\label{B.11}
\end{equation}
where $\mu$ is the chemical potential per particle in the local rest frame and $p_\mu$ the mechanical momentum per particle. Stationarity implies $\mathcal{E}=\text{const}$, hence \eqref{B.4}.

The superconducting order parameter $\psi=|\psi|e^{i\theta}$ of charge $q$ is minimally coupled to $A_\mu$. The gauge-invariant condensate 1-form is $p_\mu=\hbar\,\partial_\mu\theta - q\,A_\mu$. Projecting $p_\mu$ on the static observers gives the local Josephson relation
\begin{equation}
U^\mu p_\mu=\mu \quad\Rightarrow\quad
U^\mu\partial_\mu\theta=\frac{1}{\hbar}\,(\mu - q\,\Phi),
\label{B.12}
\end{equation}
which is \eqref{2.3} in the main text when expressed in Killing time. The spatial projection gives the superflow momentum on $\Sigma_t$.

For a static superconductor at rest with respect to $U^\mu$, the spatial superflow momentum is
\begin{equation}
p_i=\hbar\,\partial_i\theta - q A_i ,
\label{B.13}
\end{equation}
and the gauge-invariant phase drop across a weak link is the line integral of $p_i$ along any transversal (cf. Appendix \ref{apdxA}). A local change $\delta\tilde\mu$ between banks in the proper frame produces a phase rate, with respect to Killing time,
\begin{equation}
\frac{d\theta}{dt}
=\frac{\alpha}{\hbar}\,\delta\tilde\mu
=\frac{q}{\hbar}\,\alpha\,V^{\rm proper},
\label{B.14}
\end{equation}
which reproduces \eqref{2.18} after identifying $V^{\rm proper}=\delta\tilde\mu/q$. The difference of the two banks then gives the redshifted bias entry in the Josephson phase evolution, which in turn controls the scaling of bias-inferred critical currents in Section \ref{sec4}.

Note that the quantities $T$, $\mu$, $\tilde\mu$, and the combinations $\alpha T$ and $\alpha\tilde\mu$ are invariant under gauge transformations and under slice-preserving diffeomorphisms. The redshift factor $\alpha$ is a scalar constructed from the Killing field, so these relations are covariant statements of equilibrium.

One may recover \eqref{B.9} by maximizing the total entropy at fixed total energy and charge in a static spacetime. Introducing Lagrange multipliers $\beta$ and $\beta\mu_\infty$ for the constraints $\int \sqrt{\gamma}\,u/\alpha=\text{const}$ and $\int \sqrt{\gamma}\,n=\text{const}$, the stationarity condition gives $T^{-1}=\beta\,\alpha$ and $\tilde\mu=\mu_\infty/\beta\alpha$. Thus $\alpha T$ and $\alpha\tilde\mu$ are spatial constants, again reproducing \eqref{B.9}. The constants are the intensive variables “at infinity,” which are the natural control parameters for asymptotic observers.

Equations \eqref{B.1}-\eqref{B.14} collect the TE relations and their charged generalization in a form adapted to superconductors on static slices. They justify the equilibrium priors used in the main text and provide the bridge from local biases to the Killing-time phase evolution required throughout.

\section{Derivation of the \texorpdfstring{$\alpha^{2}$}{} scaling from \texorpdfstring{$j^{\mu}$}{} and Killing energy} \label{apdxC}
This appendix gives an invariant derivation of how asymptotic (Killing-time) measurements relate to local, proper-frame quantities for a short junction on a static spacetime with lapse $\alpha=\sqrt{-\xi^\mu\xi_\mu}$ for the timelike Killing field $\xi^\mu$.

For any smooth two-surface $S\subset\Sigma_t$ that threads the junction once, the conserved charge flux per unit Killing time is (cf. \eqref{2.16})
\begin{equation}
I_{\infty}[S]=\int_S \alpha\, j^\mu d\Sigma_\mu.
\label{C.1}
\end{equation}
Writing the proper flux measured by static observers as $I_{\rm proper}[S]=\int_S j^\mu d\Sigma_\mu$, then when $\alpha$ is approximately constant over the junction area,
\begin{equation}
I_{\infty}\simeq \alpha\, I_{\rm proper}.
\label{C.2}
\end{equation}
This “single $\alpha$” factor is purely kinematic and independent of microscopic constitutive details.

For a charged condensate, let $\tilde\mu\equiv\mu-q\Phi$ be the local electrochemical potential (Appendix \ref{apdxB}). The conserved Killing-energy per particle is
\begin{equation}
\mathcal{E}_\infty \equiv \alpha\,\tilde\mu,
\label{C.3}
\end{equation}
so the asymptotic (gauge-invariant) bias across the device is
\begin{equation}
V_\infty \equiv \frac{\Delta\mathcal{E}_\infty}{q}
= \frac{\Delta(\alpha\,\tilde\mu)}{q}.
\label{C.4}
\end{equation}
For a short junction with nearly uniform lapse on the banks, $\alpha_1\simeq\alpha_2\simeq\alpha$, one has
\begin{equation}
V_\infty \simeq \alpha\,V^{\rm proper}, \qquad
V^{\rm proper}\equiv \frac{\Delta\tilde\mu}{q}.
\label{C.5}
\end{equation}

The local Josephson phase-evolution law with respect to Killing time $t$ is (for an isolated junction; add the Faraday term for loop circuits)
\begin{equation}
\frac{d}{dt}\Delta\varphi
= \frac{q}{\hbar}\,\alpha\,V^{\rm proper}.
\label{C.6}
\end{equation}
Using \eqref{C.5} to trade the local drop for the asymptotic control gives the asymptotic Josephson relation
\begin{equation}
\frac{d}{dt}\Delta\varphi
= \frac{q}{\hbar}\,V_\infty,
\label{C.7}
\end{equation}
and for $q=2e$ \cite{Josephson:1962zz,Tiesinga_2021},
\begin{equation}
\omega_J \equiv \frac{d}{dt}\Delta\varphi
= \frac{2e}{\hbar}\,V_\infty.
\label{C.8}
\end{equation}
Thus, when the control parameter is the asymptotic bias $V_\infty$, the conversion from control to phase rate brings in no extra powers of $\alpha$.

The local current-phase relation reads
\begin{equation}
I_{\rm proper}=I_c^{\rm proper}\ \sin\Delta\varphi .
\label{C.9}
\end{equation}
Using the flux map \eqref{C.2}, the asymptotic current is
\begin{equation}
I_{\infty}=\alpha\,I_c^{\rm proper}\ \sin\Delta\varphi .
\label{C.9$'$}
\end{equation}
Equations \eqref{C.7}-\eqref{C.9} are the two invariant conversions needed to interpret any $I$-$\Delta\varphi$ relation under a specified asymptotic control $V_\infty$. In particular, the critical current inferred from asymptotic measurements scales as $I_{c,\infty}=\alpha\,I_c^{\rm proper}$.

Under a DC $V_\infty$, \eqref{C.8} gives a linear phase ramp; time averaging produces the usual harmonic content of $\sin\Delta\varphi$, with amplitude set by $I_{c,\infty}$ above. In loop geometries, include the Faraday term $+\,(q/\hbar)\,d(\Phi_{\rm loop}/\Phi_0)/dt$ in \eqref{C.6}-\eqref{C.7}.

The same results follow from the Killing-energy balance. The asymptotic power delivered to the device is $P_\infty=(I_\infty/q)\,\Delta\mathcal{E}_\infty$, while the proper power is $P_{\rm proper}=I_{\rm proper} V^{\rm proper}$. From \eqref{C.2}-\eqref{C.5},
\begin{equation}
P_\infty
=\frac{I_\infty}{q}\,\Delta\mathcal{E}_\infty
\simeq \frac{\alpha I_{\rm proper}}{q}\,\left(\alpha\,\Delta\tilde\mu\right)
=\alpha^{2}\,P_{\rm proper}.
\label{C.10}
\end{equation}
Thus, a given asymptotic power corresponds to a proper power larger by a factor $\alpha^{-2}$, and vice versa. This does not alter the phase-bias conversion \eqref{C.7}-\eqref{C.8}; it reflects the product structure “current $\times$ energy per charge.”

The derivation assumes a single lapse $\alpha$ across the device region and ignores $\mathcal{O}\left((L/\ell_\alpha)^2\right)$ gradients, with $L$ the junction size and $\ell_\alpha\equiv |\nabla\ln\alpha|^{-1}$. In that regime, the only redshift maps needed by an asymptotic observer are the scalar relations \eqref{C.2} and \eqref{C.3}-\eqref{C.5}, together with the Josephson kinematics \eqref{C.6}-\eqref{C.8}. Beyond this regime, one must include spatial variations of $\alpha$ and geometric corrections to both $j^\mu d\Sigma_\mu$ and the scalar $\alpha\,\tilde\mu$.

\section{Shapiro delay in Schwarzschild to leading orders and closed kernels for phase-locked drives} \label{apdxD}
We collect the propagation phases that enter the RF-driven analysis of Section \ref{sec3} and the vertical SQUID of Section \ref{sec5}. The drive is generated at infinity with angular frequency $\Omega_\infty$. In a static Schwarzschild exterior, the field is nondispersive, so the monochromatic phase at a reception point is $\Omega_\infty$ times the coordinate-time delay along the past-directed null geodesic(s) from the source to that point.

For radial rays, sending a receiver from radius $r$ to a reference radius $R\to\infty$ and subtracting the flat-space piece gives the standard Shapiro delay
\begin{equation}
\Delta t_{\rm Shapiro}^{\rm(rad)}(r)
=\frac{r_s}{c}\,\ln\,\frac{R-r_s}{r-r_s}\xrightarrow{R\to\infty}
\frac{r_s}{c}\,\ln\frac{1}{1-r_s/r}.
\label{D.4}
\end{equation}
The corresponding phase referenced to a distant clock is
\begin{equation}
\psi_{\rm rad}(r)=\Omega_\infty\,\Delta t_{\rm Shapiro}^{\rm(rad)}(r)
=\frac{\Omega_\infty r_s}{c}\,\ln\,\frac{1}{1-r_s/r}.
\label{D.5}
\end{equation}
For two receivers at $r_1$ and $r_2$,
\begin{equation}
\Delta\psi_{\rm rad}
=\psi_{\rm rad}(r_1)-\psi_{\rm rad}(r_2)
=\frac{\Omega_\infty r_s}{c}
\ln\frac{1-r_s/r_2}{1-r_s/r_1}.
\label{D.6}
\end{equation}
These are the expressions used in Secs. \ref{sec3} and \ref{sec5}.

\begin{figure}
    \centering
    \includegraphics[width=\columnwidth]{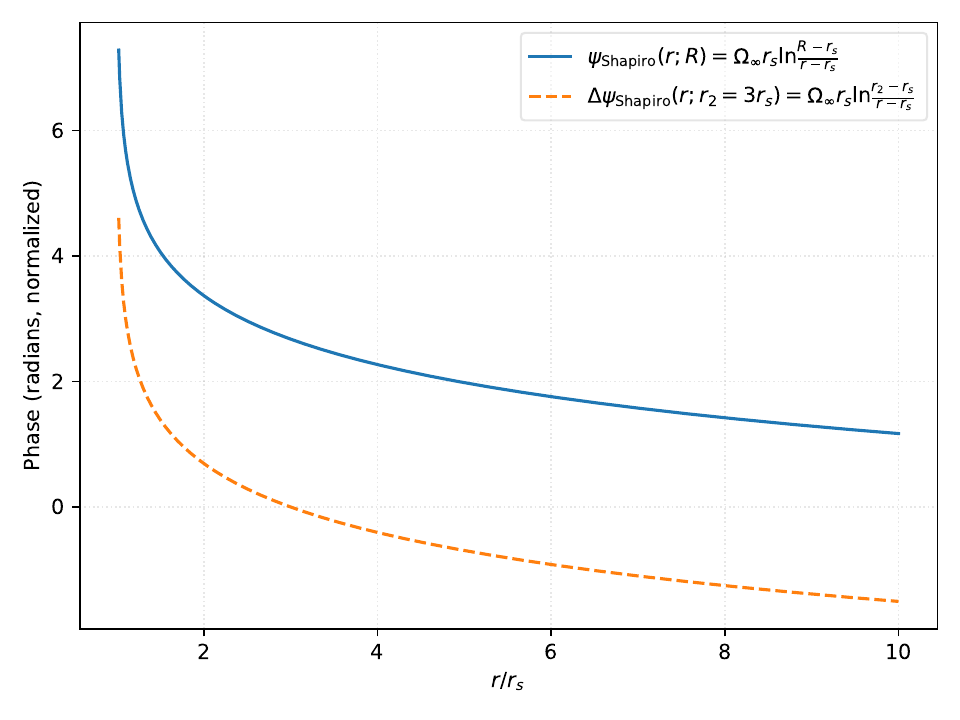}
    \caption{Propagation phase vs radius. The Shapiro contribution to the RF propagation phase from a distant source at \(R\) to radius \(r\) grows logarithmically as \(r\to r_s\): \(\psi_{\rm Shapiro}(r,R)=\Omega_\infty r_s \ln[(R-r_s)/(r-r_s)]\) (solid). The arm-to-arm phase difference after geometric path equalization depends only on the radii, \(\Delta\psi_{\rm Shapiro}(r,r_2)=\Omega_\infty r_s \ln[(r_2-r_s)/(r-r_s)]\) (dashed, with \(r_2=3r_s)\).}
    \label{fig5}
\end{figure}

The radial null propagation in Schwarzschild satisfies \(dt/dr=[1-r_s/r]^{-1}\) (see Figure \ref{fig5}). At fixed asymptotic frequency \(\Omega_\infty\), the accumulated phase is \(\psi=\Omega_\infty\int dt\). Separating the flat-space piece from the gravitational delay yields the familiar logarithmic Shapiro term. The solid curve shows that this term diverges only logarithmically at the horizon, a slow but unbounded growth that becomes relevant when integrating over many wavelengths. The dashed curve is the physically measurable quantity in a vertical SQUID once geometric path lengths are matched: the dependence on the remote source position \(R\) cancels, and the phase offset reduces to a log of radius differences. This \(\Delta\psi\) is precisely the parameter that shifts the RF interference envelope (as seen in Figure \ref{fig4}) while leaving DC lobe centers fixed.

Next is to consider non-radial rays. For impact parameter $b$ and emitter/receiver at $\boldsymbol{x}_e,\boldsymbol{x}_r$ with $r_{e,r}\gg r_s$,
\begin{align}
\Delta t_{\rm Shapiro}
&=\frac{2GM}{c^3}\,
\ln\left(\frac{r_e+r_r+R}{r_e+r_r-R}\right),\nonumber \\ 
r_{e,r}&=|\boldsymbol{x}_{e,r}|,\; R=|\boldsymbol{x}_r-\boldsymbol{x}_e|,
\label{D.7}
\end{align}
so the propagation phase is $\psi=\Omega_\infty\,\Delta t_{\rm Shapiro}$. If Euclidean path lengths are equalized by the RF network, the differential phase between receivers at $r_1,r_2$ is dominated by the Shapiro part,
\begin{equation}
\Delta\psi
=\Omega_\infty\left(\Delta t_{\rm Shapiro}(r_1)-\Delta t_{\rm Shapiro}(r_2)\right),
\label{D.8}
\end{equation}
with $\Delta t_{\rm Shapiro}$ evaluated via \eqref{D.7} (nonradial) or via \eqref{D.5}-\eqref{D.6} (radial).

For time averaging on a Shapiro step, it is convenient to assemble a closed “drive-to-phase” kernel. For a monochromatic source at infinity with complex amplitude $V_{\rm rf}^\infty$, the proper drive at a receiver $x$ is \cite{Shapiro:1964uw,Possel:2019zbu}
\begin{equation}
V^{\rm proper}(t,x)=\frac{1}{\alpha(x)}\,\mathrm{Re}\left\{\kappa(x)\,V_{\rm rf}^\infty\,e^{\,i[\Omega_\infty t-\psi(x)]}\right\},
\label{D.11}
\end{equation}
which is the local version of \eqref{3.7} with the correct $1/\alpha$ conversion from asymptotic to proper voltage. Integrating the redshifted Josephson law \eqref{2.9}/\eqref{3.2} in time then gives
\begin{equation}
\Delta\varphi(t)=\omega_0\,t
+\mathrm{Im}\left\{ \mathcal{K}(\Omega_\infty,x)\,e^{\,i\Omega_\infty t}\right\}
+\varphi_0,
\end{equation}
\begin{equation}
\mathcal{K}(\Omega_\infty,x)
=\frac{2e}{\hbar}\,\frac{\kappa(x)\,V_{\rm rf}^\infty}{i\,\Omega_\infty}\,
e^{-i\psi(x)} .
\label{D.12}
\end{equation}
Thus, the closed kernel entering the Jacobi-Anger average is the complex number
\begin{equation}
K(x,\Omega_\infty)\equiv \frac{2e}{\hbar}\,\frac{\kappa(x)\,V_{\rm rf}^\infty}{\Omega_\infty}\,
e^{-i\psi(x)}.
\label{D.13}
\end{equation}
For two receivers $x_1,x_2$, the effective differential kernel used in \eqref{3.8}-\eqref{3.13} and \eqref{5.17}-\eqref{5.18} is
\begin{equation}
K_{\rm eff}(\Omega_\infty)
=K(x_1,\Omega_\infty)-K(x_2,\Omega_\infty)
\end{equation}
\begin{equation}=\frac{2e}{\hbar}\,\frac{V_{\rm rf}^\infty}{\Omega_\infty}\,
\left[\kappa_1 e^{-i\psi_1}-\kappa_2 e^{-i\psi_2}\right],
\label{D.14}
\end{equation}
so that the modulation index and phase in the Bessel envelope are
\begin{equation}
a=\big|K_{\rm eff}\big|,
\qquad
\psi=\arg K_{\rm eff}.
\label{D.15}
\end{equation}
Equations \eqref{D.12}-\eqref{D.15} now match the main-text definitions $a=(2e/\hbar)|V_{\rm rf,eff}^{\infty}|/\Omega_\infty$ and $V_{\rm rf,eff}^{\infty}\equiv \kappa_1V_{\rm rf}^{\infty}e^{-i\psi_1}-\kappa_2V_{\rm rf}^{\infty}e^{-i\psi_2}$, i.e. no extra $\alpha$ factors appear in the RF kernel.

\begin{figure}
    \centering
    \includegraphics[width=\columnwidth]{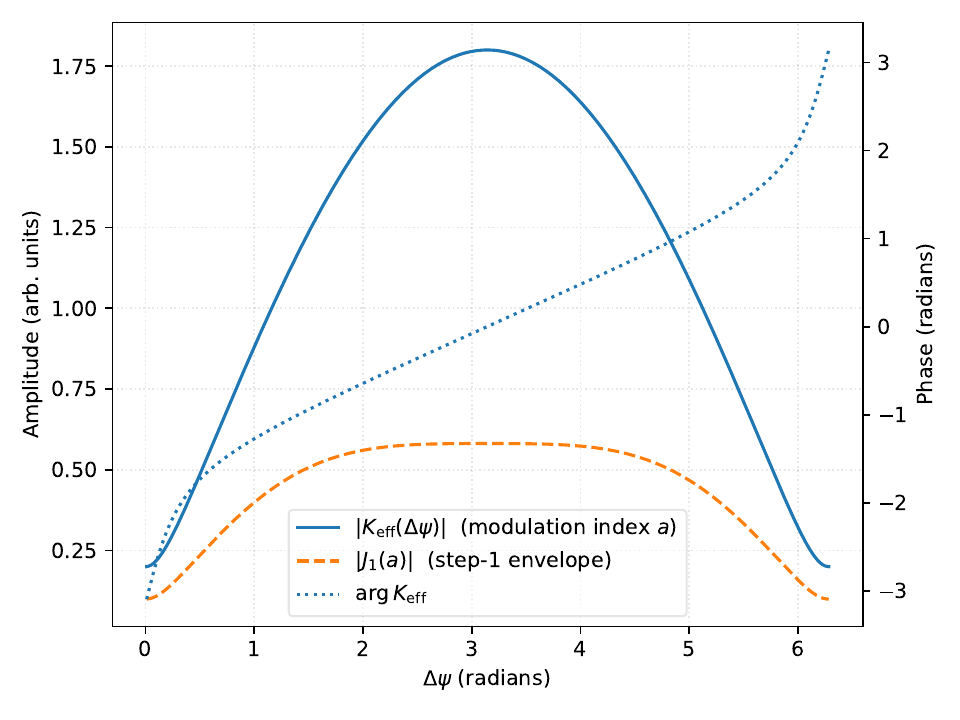}
    \caption{Kernel magnitude and phase. Effective drive-to-phase kernel \(K_{\rm eff}(\Delta\psi)=A_1 - A_2 e^{i\Delta\psi}\) for a two-arm vertical SQUID with unequal RF couplings \(A_1=0.80\), \(A_2=1.00\). The modulation index is \(a=|K_{\rm eff}|\) (solid), the step-1 envelope follows \(|J_1(a)|\) (dashed), and the effective offset phase is \(\arg K_{\rm eff}\) (dotted). Propagation-phase differences \(\Delta\psi\) therefore control both coupling strength and phase.}
    \label{fig6}
\end{figure}

Figure \ref{fig6} condenses the RF coupling mechanics into a single object, \(K_{\rm eff}\). Unequal arm drives interfere vectorially; as \(\Delta\psi\) is swept from \(0\) to \(2\pi\), the magnitude \(|K_{\rm eff}|\) ranges between \(|A_1-A_2|\) and \(A_1+A_2\). Since the Josephson phase modulation is \(a=|K_{\rm eff}|\) (in normalized units), the step-\(n\) amplitudes are governed by Bessel factors \(J_n(a)\). The dashed curve shows the \(n=1\) envelope tracking \(|J_1(a)|\): it peaks when \(a\) approaches the first maximum of \(J_1\) and collapses where \(|K_{\rm eff}|\) is minimized by destructive interference. Meanwhile, \(\arg K_{\rm eff}\) supplies the phase offset that shifts RF interference lobes (as seen in Figure \ref{fig4}). The upshot is that gravitationally induced propagation phases enter only through \(\Delta\psi\), simultaneously modulating the coupling strength and setting the phase reference for Shapiro-step envelopes, while leaving the step positions (set by asymptotic metrology) unchanged.

Note that (i) with symmetric coupling $\kappa_1=\kappa_2$ and equalized Euclidean paths, $K_{\rm eff}$ depends on gravity only through the (small) Shapiro phases $\psi_i$; redshift by itself does not generate a differential modulation, which is consistent with Sec. \ref{ssec3.2}’s use of asymptotic control variables. (ii) Lensing-induced amplitude corrections and higher-order post-Newtonian terms in $\psi$ are subleading for short devices with $L\ll R_c$ and rays far from the photon sphere; they can be incorporated by multiplying $K$ by a slowly varying real factor and augmenting $\psi$ with higher-PN terms without altering \eqref{D.13}-\eqref{D.15}. For recent theory and observation of potential-only gravitational phases and delays, see \cite{Chiao:2023ezj,Overstreet:2021hea}.

\section{Rindler near-horizon limit and proper/coordinate conversions for GL parameters} \label{apdxE}
This appendix uses asymptotic currents and voltages because its purpose is to analyze energy accounting relative to the Killing time, which is a natural notion in the near-horizon regime. For terrestrial or weak-field experiments, one should translate the results using the dictionary of Sec. \ref{ssec2.3}, replacing \(I_\infty\) and \(V_\infty\) by their proper counterparts.

Near the Schwarzschild horizon $r=r_s$ we adopt a local “Rindler patch” adapted to the static observers. Let $\rho$ denote the proper distance from the horizon along a static radial ray and let $x,y$ be orthogonal proper coordinates tangent to the horizon 2-sphere at the reference point. To leading order in $\rho/r_s$ and on scales $\ll r_s$ (so that spherical curvature is negligible across the device), the metric takes the Rindler form \cite{Wald:1999vt}
\begin{equation}
ds^2 = -(\kappa\rho)^2 c^2\, dT^2 + d\rho^2 + dx^2 + dy^2,
\quad \kappa=\frac{c^2}{4GM}=\frac{1}{2r_s},
\label{E.1}
\end{equation}
with Killing field $\xi^\mu=(\partial_T)^\mu$ and lapse $\alpha(\rho)=\kappa\rho$. Static slices $\Sigma_T$ have induced metric $\gamma_{ij}\,dx^i dx^j = d\rho^2+dx^2+dy^2$, unit normal $n^\mu=\xi^\mu/\alpha$, and volume element $\sqrt{\gamma}\,d^3x=d\rho\,dx\,dy$. The orthonormal triad aligned with $(\rho,x,y)$ coincides with the coordinate basis on $\Sigma_T$ at this order, so tensors pulled back to $\Sigma_T$ have no hidden lapse factors beyond those already explicit in the $t$-$\rho$ block of \eqref{E.1}.

The Ginzburg-Landau functional on $\Sigma_T$ is the intrinsic expression \eqref{4.1} with $\gamma_{ij}$ now flat in the $(\rho,x,y)$ coordinates of \eqref{E.1}. The order parameter $\psi=|\psi|e^{i\theta}$, effective mass $m_*$, and quartic coefficient $b>0$ are local scalars; the quadratic coefficient $a(T)$ carries the temperature dependence. The spatial gauge potential is $A_i$ pulled back to $\Sigma_T$, with covariant derivative $D_i$ as in \eqref{4.2}. In the Rindler patch, one may therefore write the kinetic term as $\gamma^{ij}(D_i\psi)^*(D_j\psi) = |D_\rho\psi|^2+|D_x\psi|^2+|D_y\psi|^2$, and all spatial inner products are Euclidean to this order. The GL coherence lengths defined locally by
\begin{equation}
\xi_{L,R}^2=\frac{\hbar^2}{2m_*|a_{L,R}|},\qquad
\xi_N^2=\frac{\hbar^2}{2m_* a_N},
\label{E.2}
\end{equation}
are proper lengths on $\Sigma_T$; no lapse enters their definition because the kinetic term is purely spatial. Likewise, a junction of thickness $L$ along its unit normal and cross-sectional area $A$ is proper geometric data on $\Sigma_T$. The local linearized weak-link problem across a barrier aligned with the unit normal $\hat{\ell}^i$ is exactly the flat-space problem in the metric $\gamma_{ij}$, so the solution quoted in \eqref{4.6} and the current-phase relation \eqref{4.8}-\eqref{4.9} hold verbatim with $\ell$ the proper distance and $\gamma^{\ell\ell}=1$ in the orthonormal frame.

Two physically distinct thermal ensembles are relevant near a horizon. If the banks are thermostatted locally at a fixed proper temperature $T$, then $a(T)$ and the coherence lengths in \eqref{E.2} are independent of $\alpha$. If instead the banks are in radiative equilibrium with a bath specified at infinity, Tolman’s law $\alpha T=\text{const}$ [cf. \eqref{B.9}] implies $T(\rho)=T_\infty/\alpha(\rho)=T_\infty/(\kappa\rho)$. As $\rho\to 0$, the local temperature increases and drives $a(T)$ toward its normal-state value; the intrinsic weak-link parameters $\xi_N\,,\xi_{L,R}$, and the interface suppression factors then acquire $\rho$-dependence through $T(\rho)$. The kinematic near-horizon scalings of Section \ref{sec4}, such as $I_{c,\infty}^{\rm(phase)}\propto \alpha$ and $I_{c,\infty}^{\rm(bias)}\propto \alpha$, remain intact; only the microscopic prefactor $I_c^{\rm proper}(T)$ changes with the chosen thermal ensemble.

Electromagnetic quantities split cleanly into temporal and spatial parts with respect to the static observers. The scalar potential is $\Phi=-U^\mu A_\mu$ with $U^\mu=n^\mu$, and the spatial vector potential on $\Sigma_T$ has components $A_i$ in the orthonormal triad $(\hat{\rho},\hat{x},\hat{y})$. The local electric and magnetic fields measured by static observers are $E_i=F_{i\mu}U^\mu$ and $B^i=\frac{1}{2}\epsilon^{ijk}F_{jk}$, with $\epsilon^{ijk}$ the Levi-Civita tensor of $\gamma_{ij}$. These definitions are gauge invariant and independent of $\alpha$. In particular, the gauge-invariant spatial momentum $p_i=\hbar\,\partial_i\theta-qA_i$ and the spatial current density $\mathcal{J}^i=(q\hbar/m_*)\,\Im(\psi^*\gamma^{ij}D_j\psi)$ [cf. \eqref{2.6} and \eqref{4.4}] are computed with the flat $\gamma_{ij}$ of \eqref{E.1} and involve no lapse factors.

The proper/coordinate conversions needed to compare with expressions written in Schwarzschild coordinates are straightforward. Let $(r,\theta,\phi)$ be the Boyer-Lindquist-like coordinates on a constant-$t$ slice near $r_s$. In a small patch centered at $(r_s,\theta_0,\phi_0)$,
\begin{equation}
\rho \simeq \frac{r-r_s}{\sqrt{1-r_s/r}}\ \xrightarrow{r\to r_s}\ 2\sqrt{r_s(r-r_s)},\end{equation}\begin{equation}
x \simeq r_s\,(\theta-\theta_0),\quad
y \simeq r_s\sin\theta_0\,(\phi-\phi_0),
\label{E.3}
\end{equation}
and the induced metric reduces to $d\rho^2+dx^2+dy^2$ at leading order, matching \eqref{E.1}. The spatial gradient and divergence map as $\nabla_\Sigma\to (\partial_\rho\,\partial_x\,\partial_y)$, while the area and volume elements become $dA=dx\,dy$ and $dV=d\rho\,dx\,dy$. Thus, a barrier whose normal is radial in Schwarzschild coordinates has proper thickness $L=\Delta \rho$, and a disk-like cross-section of coordinate radii $(\Delta\theta,\Delta\phi)$ at $(\theta_0,\phi_0)$ has proper area $A\simeq r_s^2\sin\theta_0\,\Delta\theta\,\Delta\phi$. Components of $A_i$ convert by the obvious Jacobians in \eqref{E.3}, but no lapse enters because $A_i$ is intrinsic to $\Sigma_T$.

With these conventions, the near-horizon limits quoted in the main text follow transparently. The redshift factor is $\alpha(\rho)=\kappa\rho$. The local GL solution on $\Sigma_T$ yields a proper current-phase law $I_{\rm proper}=I_c^{\rm proper}\sin\Delta\varphi$ with $I_c^{\rm proper}$ determined by $L, A, \xi_N, \xi_{L,R}$, and interface parameters, all understood as proper quantities at the junction’s location. The mapping to asymptotic observables uses only the hypersurface flux rule and the bias-phase conversion developed in Section \ref{sec2}: per unit Killing time,
\begin{equation}
I_{\infty}\simeq \alpha\,I_{\rm proper},\qquad
\frac{d\theta}{dT}=\frac{q}{\hbar}\,\alpha\,V^{\rm proper},
\label{E.4}
\end{equation}
which reproduce, in the Rindler gauge, the phase-clamped scaling $I_{c,\infty}^{\rm(phase)}\sim \alpha I_c^{\rm proper}$ and the bias-inferred scaling $I_{c,\infty}^{\rm(bias)}\sim \alpha I_c^{\rm proper}$ [cf. \eqref{4.12}-\eqref{4.17} and \eqref{4.19}-\eqref{4.20}]. Nothing in \eqref{E.4} depends on the microscopic GL parameters beyond their appearance in $I_c^{\rm proper}$.

Finally, we record the domain of validity. The Rindler reduction \eqref{E.1} is accurate when the device size $L$ and the loop dimensions $\sqrt{A}$ satisfy $L\,\sqrt{A}\ll r_s$, and when the lapse varies little across the junction, $L\ll \ell_\alpha\equiv \alpha/|\nabla\alpha|=\rho$ for Schwarzschild at leading order. Geometric corrections scale as $(L/\rho)^2$ from lapse inhomogeneity and as $(L/R_c)^2$ from background curvature on $\Sigma_T$, with $R_c\sim r_s/\sqrt{48}$ near the horizon. Within these bounds, the only near-horizon modifications to DC transport and RF phase locking are the explicit $\alpha$-factors associated with the Killing projection and, optionally, the Tolman rescaling of the GL coefficients when the thermal ensemble is fixed at infinity rather than locally.

\bibliography{ref}

\end{document}